\documentclass[acmlarge]{acmart}

\usepackage{multirow}
\usepackage{amssymb}
\usepackage{pifont}
\usepackage{amsmath}
\usepackage{diagbox}
\usepackage{url}
\usepackage{hyperref}
\usepackage{longtable}
\hypersetup{
    colorlinks=true,
    linkcolor=blue,
    filecolor=magenta,      
    urlcolor=cyan,
}
\allowdisplaybreaks

\DeclareMathOperator*{\argmax}{arg\,max}
\DeclareMathOperator*{\argmin}{arg\,min}
\newtheorem{definition}{Definition}
\newtheorem{assumption}{Assumption}[section]
\newcommand{\independent}{\protect\mathpalette{\protect\independenT}{\perp}}
\def\independenT#1#2{\mathrel{\rlap{$#1#2$}\mkern2mu{#1#2}}}

\setcopyright{none}
\renewcommand\footnotetextcopyrightpermission[1]{}
\begin{document}

\title{A Survey on Causal Inference}

\author{Liuyi Yao}
\affiliation{%
  \institution{University at Buffalo, USA}
  \city{Buffalo}
  \state{New York}
}
\email{liuyiyao@buffalo.edu}

\author{Zhixuan Chu}
\affiliation{%
  \institution{University of Georgia, USA}
  \city{Athens}
  \state{Georgia}
  }
\email{zhixuan.chu@uga.edu}

\author{Sheng Li}
\affiliation{%
  \institution{University of Georgia, USA}
  \city{Athens}
  \state{Georgia}
  }
\email{sheng.li@uga.edu}

\author{Yaliang Li}
\affiliation{%
  \institution{Alibaba Group, USA}
  \city{Bellevue}
  \state{Washington}
}
\email{yaliang.li@alibaba-inc.com}

\author{Jing Gao}
\affiliation{%
  \institution{University at Buffalo, USA}
  \city{Buffalo}
  \state{New York}
}
\email{jing@buffalo.edu}

\author{Aidong Zhang}
\affiliation{%
  \institution{University of Virginia, USA}
  \city{Charlottesville}
  \state{Virginia}
  }
\email{aidong@virginia.edu}

\begin{abstract}
Causal inference is a critical research topic across many domains, such as statistics, computer science, education, public policy and economics, for decades. Nowadays, estimating causal effect from observational data has become an appealing research direction owing to the large amount of available data and low budget requirement, compared with randomized controlled trials. Embraced with the rapidly developed machine learning area, various causal effect estimation methods for observational data have sprung up. In this survey, we provide a comprehensive review of causal inference methods under the potential outcome framework, one of the well known causal inference framework. The methods are divided into two categories depending on whether they require all three assumptions of the potential outcome framework or not. For each category, both the traditional statistical methods and the recent machine learning enhanced methods are discussed and compared. The plausible applications of these methods are also presented, including the applications in advertising, recommendation, medicine and so on. Moreover, the commonly used benchmark datasets as well as the open-source codes are also summarized, which facilitate researchers and practitioners to explore, evaluate and apply the causal inference methods.  
\end{abstract}

\maketitle
\fancyfoot{}

\section{Introduction}

In everyday language, correlation and causality are commonly used interchangeably, although they have quite different interpretations. Correlation indicates a general relationship: two variables are correlated when they display an increasing or decreasing trend~\cite{altman2015points}. Causality is also referred to as cause and effect where the cause is partly responsible for the effect, and the effect is partly dependent on the cause. Causal inference is the process of drawing a conclusion about a causal connection based on the conditions of the occurrence of an effect. The main difference between causal inference and inference of correlation is that the former analyzes the response of the effect variable when the cause is changed~\cite{pearl2009causal, stephen2007counterfactuals}. 

It is well known that ``\textit{correlation does not imply causation}.'' For example, a study showed that girls have breakfast normally have lightweight than the girls who don't, and thus concluded that having breakfast can help to lose weight. But in fact, these two events may just have correlation instead of causality. Maybe the girls who have breakfast everyday have a better lifestyle, such as exercise frequently, sleep regularly, and have a healthy diet, which finally makes them have lightweight. In this case, having a better lifestyle is the common cause of both having breakfast and lightweight, and thus we also can treat it as a confounder of the causality between having breakfast and lightweight.

In many cases, it seems obvious that one action can cause another; however, there exists also many cases that we cannot easily tease out and make sure the relationship. Therefore, learning causality is one dauntingly challenging problem. The most effective way of inferring causality is to conduct a randomized controlled trial, which randomly assigns participants into a treatment group or a control group. As the randomized study is conducted, the only expected difference between the control and treatment groups is the outcome variable being studied. However, in reality, randomized controlled trials are always time-consuming and expensive, and thus the study cannot involve many subjects, which may be not representative of the real-world population a treatment/intervention would eventually target. Another issue is that the randomized controlled trials only focus on the average of samples, and it doesn't explain the mechanism or pertain for individual subjects. In addition, ethical issues also need to be considered in most of the randomized controlled trials, which largely limits its applications. 
Therefore, instead of the randomized controlled trials, the observational data is a tempting shortcut. Observational data is obtained by the researcher simply observing the subjects without any interfering. That means, the researchers have no control over treatments and subjects, and they just observe the subjects and record data based on their observations. From the observational data, we can find their actions, outcomes, and information about what has occurred, but cannot figure out the mechanism why they took a specific action. For the observational data, the core question is how to get the counterfactual outcome. For example, we want to answer this question "would this patient have different results if he received a different medication?" Answering such counterfactual questions is challenging due to two reasons~\cite{schwab2019learning}: the first one is that we only observe the factual outcome and never the counterfactual outcomes that would potentially have happened if they have chosen a different treatment option. The second one is that treatments are typically not assigned at random in observational data, which may lead the treated population differs significantly from the general population.

To solve these problems in causal inference from observational data, researchers develop various frameworks, including the potential outcome framework ~\cite{splawa1990application, rubin1974estimating} and the structural causal model ~\cite{pearl1995causal,pearl2014probabilistic, pearl2009causality}. The potential outcome framework is also known as the Neyman-Rubin Potential Outcomes or the Rubin Causal Model. In the example we mentioned above, a girl would have a particular weight if she had breakfast normally everyday, whereas she would have a different weight if she didn't have breakfast normally. To measure the causal effect of having breakfast normally for a girl, we need to compare the outcomes for the same person under both situations. Obviously, it is impossible to see both potential outcomes at the same time, and one of the potential outcomes is always missing. The potential outcome framework aims to estimate such potential outcomes and then calculate the treatment effect. Therefore, the treatment effect estimation is one of the central problems in causal inference under the potential outcome framework. Another influential framework in causal inference is the structural causal model (SCM), which includes the causal graph and the structural equations. The structural causal model describes the causal mechanisms of a system where a set of variables and the causal relationship among them are modeled by a set of simultaneous structural equations. 

Causal inference has a close relationship with the machine learning area. In recent years, the magnificent bloom of the machine learning area enhances the development of the causal inference area. Powerful machine learning methods such as decision tree, ensemble methods, deep neural network, are applied to estimate the potential outcome more accurately. In addition to the amelioration on the outcome estimation model, machine learning methods also provide a new aspect to handle the confounders. Benefit from the recently deep representation learning methods, such as generative adversarial neural network, the confounder variables are adjusted by learning the balanced representation for all covariates, so that conditioning on the learned representation, the treatment assignment is independent of the confounder variables. In machine learning, the more data the better. However, in causal inference, the more data alone is not yet enough. Having more data only helps to get more precise estimates, but it cannot make sure these estimates are correct and unbiased. Machine learning methods enhance the development of causal inference, meanwhile, causal inference also helps machine learning methods.
The simple pursuit of predictive accuracy is insufficient for modern machine learning research, and correctness and interpretability are also the targets of machine learning methods. Causal inference is starting to help to improve machine learning, such as recommender systems or reinforcement learning.

In this article, we provide a comprehensive review of the causal inference methods under the potential outcome framework. We first introduce the basic concepts of the potential outcome framework as well as its three critical assumptions to identify the causal effect. After that, various causal inference methods with these three assumptions are discussed in detail, including re-weighting methods, stratification methods, matching based methods, tree-based methods, representation-based methods, multi-task learning based methods, and meta-learning methods. Additionally, causal effect estimation methods that relax the three assumptions are also described to fulfill the needs in different settings. After introducing various causal effect estimation methods, the real-world applications that the discussed methods have great potential to benefit are discussed, including the advertisement area, recommendation area, medicine area, and reinforcement learning area as the representative examples. 

To the best of our knowledge, this is the first paper that provides a comprehensive survey for causal inference methods under the potential outcome framework. There also exist several surveys that discuss one category of the causal effect estimation methods, such as the survey of matching based methods~\cite{stuart2010matching_review}, survey of tree-based and ensemble-based method~\cite{athey2015machine}, and the review of dynamic treatment regimes~\cite{chakraborty2014dynamic}. For the structural causal model, it is suggested to refer the survey~\cite{pearl2009causal} or the book~\cite{pearl2000causality}. We will also briefly discuss the relation and difference between the two causality frameworks at the end of our survey. There is also a survey about learning causality from observational data~\cite{guo2018survey} whose content ranges from inferring the causal graph from observational data, structural causal model, potential outcome framework and their connection to machine learning. Compared with the surveys mentioned above, this survey paper mainly focuses on the theoretical background of the potential outcome framework, the representative methods across the statistic domain and machine learning domain, and how this framework and the machine learning area enhance each other. 

To summarize, our contributions of this survey are as follows:
\begin{itemize}
    \item \emph{New taxonomy:} We separate various causal inference methods into two major categories based on whether they require the three assumptions of the potential outcome framework. The category requiring three assumptions are further divided into seven sub-categories based on the way to handle the confounder variables. 
    \item \emph{Comprehensive review:} We provide a comprehensive survey of the causal inference methods under the potential outcome framework. In each category, the detailed descriptions of the representative methods, the connection and comparison between the mentioned methods, and the general summation are provided.
    \item \emph{Abundant resources:} In this survey, we list the state-of-art methods, the benchmark data sets, open-source codes, and representative applications.
\end{itemize}

The rest of the paper is organized as follows. In section~\ref{sec2}, the background of the potential outcome framework is introduced, including the basic definitions, the assumptions, and the fundamental problems with their general solutions. In Section~\ref{Section-3}, the methods under three assumptions are presented. Then, in Section~\ref{section: relax three assumptions}, we discuss the problem when some assumptions are not satisfied, and describe the methods that relax those assumptions. Next, we provide experimental guidelines in Section~\ref{sec:experiment}. Afterward, in Section~\ref{sec: application}, the typical applications of causal inference are illustrated. Finally, Section~\ref{sec: conclusions} summarizes the paper. 

\section{Basic of Causal Inference}
\label{sec2}

In this section, we present the background knowledge of causal inference, including task description, mathematical notions, assumptions, challenges and general solutions. We also give an illustrative example that will be used throughout this survey.

Generally speaking, the task of causal inference is to estimate the outcome changes if another treatment had been applied. For example, suppose there are two treatments that can be applied to patients: Medicine A and Medicine B. When applying Medicine A to the interested patient cohort, the recovery rate is $70\%$, while applying Medicine B to the same cohort, the recovery rate is $90\%$. The change of recovery rate is the effect that treatment (i.e., medicine in this example) asserts on the recovery rate. 

The above example describes an ideal situation to measure the treatment effect: applying different treatments to the same cohort. In real-world scenarios, this ideal situation can only be approximated by a randomized experiment, in which the treatment assignment is controlled, such as a completely random assignment. In this way, the group receives a specific treatment can be viewed as an approximation to the cohort we are interested in.  

However, performing randomized experiments are expensive, time-consuming, and sometimes even unethical. Therefore, estimating the treatment effect from observational data has attracted growing attention due to the wide availability of observational data. Observational data usually contains a group of individuals taken different treatments, their corresponding outcomes, and possibly more information, but \textit{without direct access to the reason/mechanism why they took the specific treatment}. Such observational data enable researchers to investigate the fundamental problem of learning the causal effect of a certain treatment without performing randomized experiments. To better introduce various treatment effect estimation methods, the following section introduces several definitions including unit, treatment, outcome, treatment effect, and other information (pre- and post-treatment variables) provided by observational data.

\subsection{Definitions}
\label{subsec:definition}
Here we define the notations under the potential outcome framework~\cite{rubin1974estimating,splawa1990application}, which is logically equivalent to another framework, the structural causal model framework \cite{judea_PO}. The foundation of potential outcome framework is that the causality is tied to treatment (or action, manipulation, intervention), applied to a unit~\cite{imbens2015causal_book}. The treatment effect is obtained by comparing units' potential outcomes of treatments. In the following, we first introduce three essential concepts in causal inference: unit, treatment, and outcome.

\begin{definition}
\textit{Unit.} A unit is the atomic research object in the treatment effect study. 
\end{definition}
A unit can be a physical object, a firm, a patient, an individual person, or a collection of objects or persons, such as a classroom or a market, at a particular time point ~\cite{imbens2015causal_book}. Under the potential outcome framework, the atomic research objects at different time points are different units. One unit in the dataset is a sample of the whole population, so in this survey, the term ``sample'' and ``unit'' are used interchangeably.

\begin{definition}
\textit{Treatment.} Treatment refers to the action that applies (exposes, or subjects) to a unit.
\end{definition}
Let $W$ ($W\in \left\{0,1,2,\dots, N_W \right\}$) denote the treatment, where $N_W + 1$ is the total number of possible treatments. In the aforementioned medicine example, Medicine A is a treatment. Most of the literatures consider the binary treatment, and in this case, the group of units applied with treatment $W = 1$ is the \textit{treated group}, and the group of units with $W = 0$ is the \textit{control group}.

\begin{definition}
\textit{Potential outcome.} For each unit-treatment pair, the outcome of that treatment when applied on that unit is the potential outcome~\cite{imbens2015causal_book}. 
\end{definition}
The potential outcome of treatment with value $w$ is denoted as $Y(W = w)$.

\begin{definition}
\textit{Observed outcome.} The observed outcome is the outcome of the treatment that is actually applied.
\end{definition}
 The observed outcome is also called factual outcome, and we use $Y^F$ to denote it where F stands for ``factual''. The relation between the potential outcome and the observed outcome is: $Y^F = Y(W = w)$ where $w$ is the treatment actually applied.

\begin{definition}
\textit{Counterfactual outcome:} Counterfactual outcome is the outcome if the unit had taken another treatment.
\end{definition}
The counterfactual outcomes are the potential outcomes of the treatments except the one actually taken by the unit. Since a unit can only take one treatment, only one potential outcome can be observed, and the remaining unobserved potential outcomes are the counterfactual outcome. In the multiple treatment case, let $Y^{CF}(W = w^{'})$ denote the counterfactual outcome of treatment with value $w^{'}$. In the binary treatment case, for notation simplicity, we use $Y^{CF}$ to denote the counterfactual outcome, and $Y^{CF} = Y(W = 1-w)$, where $w$ is the treatment actually taken by the unit.

In the observational data, besides the chosen treatments and the observed outcome, the units' other information is also recorded, and they can be separated as pre-treatment variables and the post-treatment variables. 

\begin{definition}
\textit{Pre-treatment variables:} Pre-treatment variables are the variables that will not be affected by the treatment.
\end{definition}
Pre-treatment variables are also named as \textit{background variables}, and they can be patients' demographics, medical history, and etc. Let $X$ denote the pre-treatment variables. 

\begin{definition}
\textit{post-treatment variables:} The post-treatment variables are the variables that are affected by the treatment.
\end{definition}
One example of post-treatment variables is the intermediate outcome, such as the lab test after taking the medicine in the aforementioned medicine example.

In the following sections, the terminology \emph{variable} refers to the pre-treatment variable unless otherwise specified.

\textbf{Treatment Effect}. After introducing the observational data and the key terminologies, the treatment effect can be quantitatively defined using the above definitions. The treatment effect can be measured at the population, treated group, subgroup, and individual levels. To make these definitions clear, here we define the treatment effect under binary treatment, and it can be extended to multiple treatments by comparing their the potential outcomes.  

At the population level, the treatment effect is named as the Average Treatment Effect (ATE), which is defined as: 
\begin{equation}
    \text{ATE}  = \mathbb{E}[\mathbf{Y}(W = 1) - \mathbf{Y}(W = 0)],
\end{equation}
where $\mathbf{Y}(W = 1)$ and $\mathbf{Y}(W = 0)$ are the potential treated and control outcome of the whole population respectively.

For the treated group, the treatment effect is named as Average Treatment effect on the Treated group (ATT), and it is defined as:
\begin{equation}
    \text{ATT}  = \mathbb{E}[\mathbf{Y}(W = 1)| W = 1] - \mathbb{E}[\mathbf{Y}(W = 0)|W = 1 ],
\end{equation}
where $\mathbf{Y}(W = 1)| W = 1$ and  $\mathbf{Y}(W = 0)| W = 1$ are the potential treated and control outcome of the treated group respectively.

At the subgroup level, the treatment effect is called Conditional Average Treatment Effect (CATE), which is defined as:
\begin{equation}
    \text{CATE}  = \mathbb{E}[\mathbf{Y}(W = 1)|X = x] - \mathbb{E}[\mathbf{Y}(W = 0)|X = x],
\end{equation}
where $\mathbf{Y}(W = 1)|X = x$ and  $\mathbf{Y}(W = 0)|X = x$ are the potential treated and control outcome of the subgroup with $X = x$, respectively. CATE is a common treatment effect measurement under the case where the treatment effect varies across different subgroups, which is also known as the heterogeneous treatment effect.

At the individual level, the treatment effect is called Individual Treatment Effect (ITE), and the ITE of unit $i$ is defined as:
\begin{equation}
    \text{ITE}_i = Y_i(W = 1) - Y_i(W = 0),
    \label{eqn: ITE}
\end{equation}
where $Y_i(W = 1)$ and $Y_i(W = 0)$ are the potential treated and control outcome of unit $i$ respectively. In some literatures~\cite{johansson2016learning,shalit2017estimating}, the ITE is viewed equivalent to the CATE.

\textbf{Objective}.
For causal inference, our objective is to estimate the treatment effects from the observational data. Formally speaking, given the observational dataset, $\left\{ X_i, W_i, Y_i^F \right\}_{i = 1}^{N}$, where $N$ is the total number of units in the datasets, the goal of the causal inference task is to estimate the treatment effects defined above.

\subsection{An Illustrative Example}
To better illustrate causal inference, we use the following example combined with the notations defined above to give an overview. In this example, we want to evaluate the treatment effects of several different medications for one disease, by exploiting the observational data (i.e., the electronic health records) that include demographic information of patients, the specific medication with the specific dosage taken by patients, and the outcome of medical tests. Obviously, we can only get one factual outcome for a specific patient from electronic health records, and thus the core task is to predict what would have happened if a patient took another treatment (i.e., a different medication or the same medication with a different dosage). Answering such counterfactual questions is very challenging. Therefore, we want to use causal inference to predict all of the potential outcomes for each patient over all of the medications with different dosages. Then, we can reasonably and accurately evaluate and compare the treatment effect of different medications for this disease. 

One particular point to keep in mind is that for each medication, they may have different dosages. For example, for medication A, the dosage range can be a continuous variable in the range $[a, b]$ while for medication B, the dosage can be a categorical variable that has several specific dosage regimens. 

In the aforementioned example, the units are the patients with the studied disease. 
The treatments refer to the different medications with specific dosages for this disease, and we use $W$ ($W\in \left\{0,1,2,\dots, N_W \right\}$) to denote these treatments. For example, $W_i=1$ can represent the medication $A$ with a specific dosage is taken by the unit $i$, and $W_i=2$ represents the medication $B$ with a specific dosage is taken by the unit $i$.
$Y$ is the outcome, such as one type of blood test that can measure the medication's ability to destroy the disease and lead to the recovery of the patients. Let $Y_i(W = 1)$ denote the potential outcome of medication $A$ with a specific dosage on patient $i$. 
The features of patients may include age, gender, clinical presentation, and some other medical tests, etc. Among these features, age, gender and other demographic information are pre-treatment variables that cannot be affected by taking a treatment.
Some clinical presentations and medical tests are affected by taking medications, and they are post-treatment variables.
In this example, our goal to estimate the treatment effects of different medications for this disease based on the provided observational data.

In the following sections, we will continuously use this example to explain more concepts and illustrate intuitions behind various causal inference methods.

\subsection{Assumptions}
\label{subsection: Assumptions}
In order to estimate the treatment effect, the following assumptions are commonly used in the causal inference literature.

\begin{assumption} \textbf{Stable Unit Treatment Value Assumption (SUTVA)}. The potential outcomes for any unit do not vary with the treatment assigned to other units, and, for each unit, there are no different forms or versions of each treatment level, which lead to different potential outcomes.
\label{asp: SUTVA}
\end{assumption}
This assumption emphasizes two points: The first point is the independence of each unit, that is, there are no interactions between units. In the context of the above illustrative example, one patient's outcome will not affect other patients' outcomes. 

The second point is the single version for each treatment. In the above example, Medicine A with different dosages are different treatments under the SUTVA assumption.

\begin{assumption} \textbf{Ignorability.}
Given the background variable, $X$, treatment assignment $W$ is independent to the potential outcomes, i.e., 
$W \independent {Y(W = 0), Y(W = 1)} | X$. 
\end{assumption}
In the context of the illustrative example, this ignorability assumption indicates two folds: First, if two patients have the same background variable $X$, their potential outcomes should be the same whatever the treatment assignment is, i.e., $p(Y_i(0), Y_i(1)| X =x, W = W_i) = p(Y_j(0), Y_j(1)| X = x, W = W_j)$. 
Analogously, if two patients have the same background variable value, their treatment assignment mechanism should be same whatever the value of potential outcomes they have, i.e., $p(W| X =x, Y_i(0), Y_i(1)) = p(W| X =x, Y_j(0), Y_j(1))$.
The ignorability assumption is also named as unconfoundedness assumption. 
With this unconfoundedness assumption, for the units with the same background variable $X$, their treatment assignment can be viewed as random.

\begin{assumption}
\textbf{Positivity}. For any value of $\,X$, treatment assignment is not deterministic:
\begin{equation}
   P(W = w | X = x) > 0, \quad \forall \, w \text{ and } x.
\end{equation}
\end{assumption}
If for some values of $X$, the treatment assignment is deterministic; then for these values, the outcomes of at least one treatment could never be observed. In this case, it would be unable and meaningless to estimate the treatment effect. To be more specific, suppose there are two treatments: Medicine A and Medicine B. Let's assume that patients with age greater than $60$ are always assigned with medicine A, then it will be unable and meaningless study the outcome of medicine B on those patients. In other words, the positivity assumption indicates the variability, which is important for treatment effect estimation.

In~\cite{imbens2015causal_book}, the ignorability and the positivity assumptions together are called \textit{Strong Ignorability} or \textit{Strongly Ignorable Treatment Assignment}.

With these assumptions, the relationship between the observed outcome and the potential outcome can be rewritten as:
\begin{equation}
\begin{split}
    \mathbb{E}[Y(W = w)|X = x] 
    &=\mathbb{E}[Y(W = w)|W = w, X = x] \text{ (Ignorability)}\\
    &= \mathbb{E}[Y^F|W = w, X = x],
    \label{Eqn: observed and potential outcome}
\end{split}
\end{equation}
where $Y^F$ is the random variable of the observed outcome, and $Y(W = w)$ is the random variable of the potential outcome of treatment $w$. If we are interested in the potential outcome of one specific group (either the subgroup, the treated group, or the whole population), the potential outcome can be obtained by taking expectation of the observed outcome over that group.

With the above equation, we can rewrite the treatment effect defined in Section \ref{subsec:definition} as follows:
\begin{equation}
\begin{split}
    \text{ITE}_i &= W_i Y_i^F - W_i Y_i^{CF} + (1-W_i) Y_i^{CF} - (1-W_i) Y_i^{F}\\
    \text{ATE} &= \mathbb{E}_{X}\left[\mathbb{E}[Y^F|W = 1, X = x] -\mathbb{E}[Y^F|W = 0, X = x]\right]\\
    &= \frac{1}{N}\sum_{i} \left(Y_i(W = 1) -Y_i(W = 0)\right) = \frac{1}{N}\sum_{i} \text{ITE}_i\\
    \text{ATT} &= \mathbb{E}_{\mathcal{X}_T}\left[\mathbb{E}[Y^F|W = 1, X = x] -\mathbb{E}[Y^F|W = 0, X = x]\right]\\
    & = \frac{1}{N_T}\sum_{\{i: W_i = 1\}} \left(Y_i(W = 1) -Y_i(W = 0)\right) =\frac{1}{N_T}\sum_{\{i: W_i = 1\}} \text{ITE}_i\\
    \text{CATE} &= \mathbb{E}[Y^F|W = 1, X = x] -\mathbb{E}[Y^F|W = 0, X = x]\\
    & = \frac{1}{N_{x}}\sum_{\{i: X_i = x\}} \left(Y_i(W = 1) -Y_i(W = 0)\right) = \frac{1}{N_{x}}\sum_{\{i: X_i = x\}}\text{ITE}_i
\end{split}
\label{eqn: empirical estimation of effect}
\end{equation}
where $Y_i(W = 1)$ and $Y_i(W = 0)$ are the potential treated/control outcomes of unit $i$, $N$ is the total number of units in the whole population, $N_T$ is the number of units in the treated group, and $N_x$ is the number of units in the group with $X = x$. The second line in the ATE, ATT and CATE equations are their empirical estimations. Empirically, the ATE can be estimated as the average of ITE on the entire population. Similarly, ATT and CATE can be estimated as the average of ITE on the treated group and specific subgroup separately. 

However, due to the fact that the potential treated/control outcomes can never be observed simultaneously, the key point in the treatment effect estimation is how to estimate the counterfactual outcome in ITE estimation or how to estimate the $\frac{1}{N_{*}}\sum_{i} Y_i(W = 1)$ and $\frac{1}{N_{*}}\sum_{i} Y_i(W = 0)$, where $N_{*}$ denotes $N$, $N_T$ or $N_x$. In the following section, we will discuss the challenges in estimation these terms and briefly introduce the general solutions.

\subsection{Confounders and General Solutions}
\label{genSolution}

As mentioned above, how to estimate the average potential treated/control outcome over a specific group is the core of causal inference. 
Let's take ATE as a case study: When estimating the ATE, a natural idea is to directly use the average of observed treated/control outcomes, i.e., $\hat{\text{ATE}} = \frac{1}{N_T}\sum_{i = 1}^{N_T}Y_i^F - \frac{1}{N_C}\sum_{i = 1}^{N_C}Y_j^F$, where $N_T$ and ${N_C}$ is the number of units in the treated and control group, respectively. However, due to the existence of \textit{confounders}, there is a serious problem in this estimation: this calculated ATE includes a spurious effect brought by the confounders. 

\begin{definition}
\textit{Confounders.} Confounders are the variables that affect both the treatment assignment and the outcome.
\end{definition}
Confounders are some special pre-treatment variables, such as age in the medicine example. When directly using the average of observed treated/control outcome, the calculated ATE not only includes the effect of treatment on the outcome, but also includes the effect of confounders on the outcome, which leads to the \textbf{spurious effect}.
For example, in the medicine example, age is a confounder. Age affects the recovery rate: in general, young patients have better chance to recover compared to older patients. Age also affects the treatment choice: young patients may prefer to take medicine A while older patients prefer medicine B, or for the same medicine, young patients have a different dosage with elder patients. The observational data is shown in Table~\ref{tab: Simpson's paradox}, and let's estimate ATE according to the above equation: $\hat{\text{ATE}} = \frac{1}{N_A}\sum_{i = 1}^{N_A}Y_i^F - \frac{1}{N_B}\sum_{i = 1}^{N_B}Y_j^F = 295/350 - 273/350 = 5\%$, where ${N_A}$ and ${N_B}$ is the number of patients taking Medicine A and B, respectively. 
However, we cannot conclude that Treatment A is more effective than Treatment B, because the high average recovery rate of the group taking Treatment A may be caused by the fact that most patients of this group ($270$ out of $350$) are young patients. Thus the effect of age on the recovery rate is the spurious effect, as it is mistakenly counted into the effect of treatment on the outcome. 

\begin{table}[ht]
    \centering
    \begin{tabular}{|c|c|c|}
    \toprule
    \diagbox[]{Age}{Recovery Rate}{Treatment} & Treatment A & Treatment B \\
    \midrule
     Young    & $234/270 = 87\%$ & $81/87 = \mathbf{92\%}$ \\
     \midrule
     Older & $55/80 = 69\%$ & $192/263 = \mathbf{73\%}$\\
     \bottomrule
     Overall & $289/350 = \mathbf{83\%}$ & $273/350 = 78\%$\\
     \bottomrule
    \end{tabular}
    \caption{An example to show the spurious effect of confounder variable \emph{Age}.}
    \label{tab: Simpson's paradox}
\end{table}

From Table~\ref{tab: Simpson's paradox}, we can observe another interesting phenomenon, \textit{Simpson's paradox} (or Simpson's reversal, Yule-Simpson effect, amalgamation paradox, reversal paradox)~\cite{blyth1972simpson, good1987amalgamation}, brought by the confounder. It can be observed that: in both Young and Older patient groups, Medicine B has a higher recovery rate than Medicine A; but when combining these two groups, Medicine A is the one with a higher recovery rate. This paradox is caused by the confounder variable: When compare the recovery rate in the whole group, most of the people taking medicine A are young, and the comparison shown in the table fails to eliminate the effect of age on the recovery rate. 

In addition to the spurious effect in treatment effect estimation, confounders also cause problems in counterfactual outcome estimation. As shown in Eqn.~(\ref{eqn: empirical estimation of effect}), counterfactual outcome estimation is an alternative way to estimate the ATE. Confounder variables cause selection bias, which makes the counterfactual outcome estimation more difficult.

\begin{figure}
    \centering
    \includegraphics[width=0.5\textwidth]{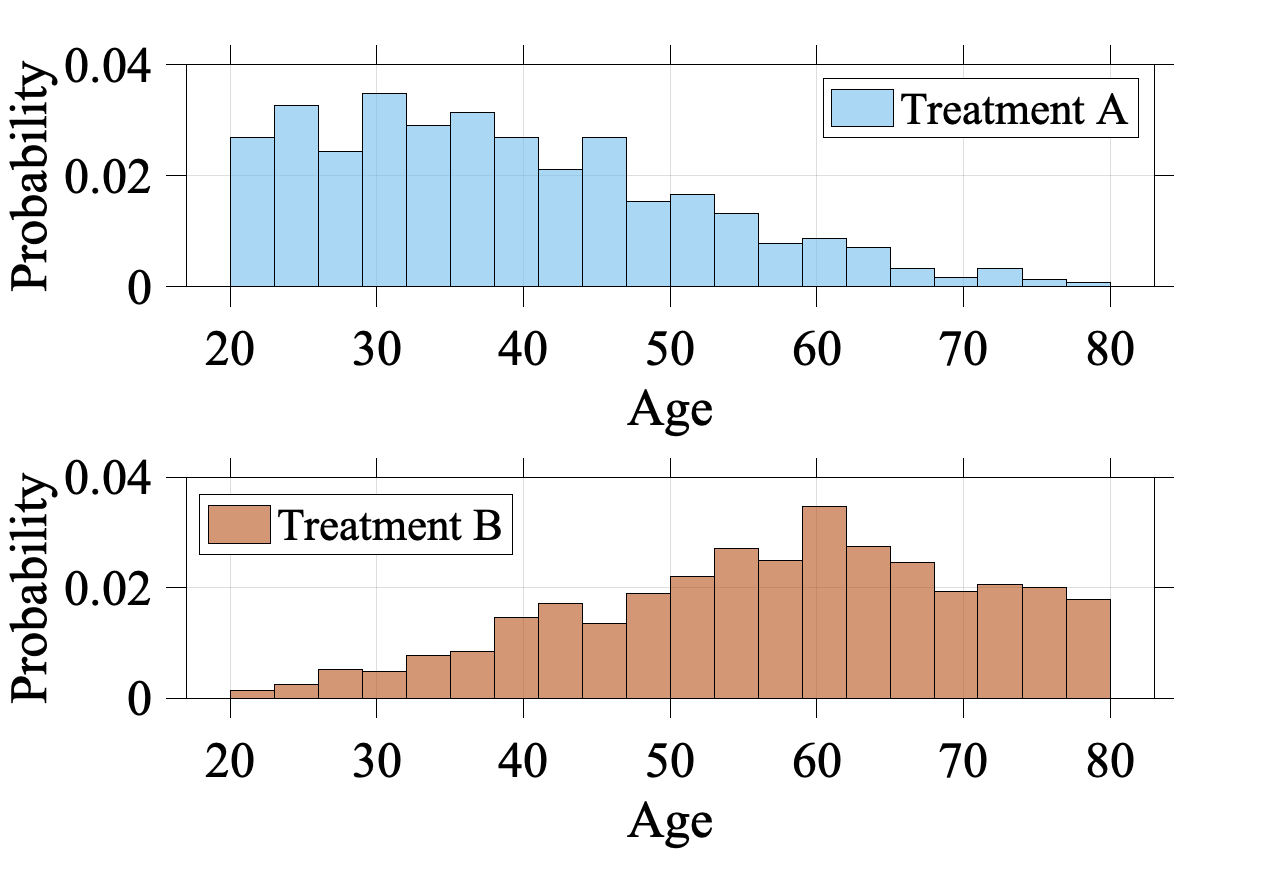}
    \caption{An example to show the selection bias caused by confounder variable \emph{Age}.}
    \label{fig: confounder}
\end{figure}

\textbf{Selection bias} is the phenomenon that the distribution of the observed group is not representative to the group we are interested in, i.e., $p(X_{obs}) \neq p(X_{*})$, where $p(X_{obs})$ and $p(X_{*})$ are the distributions of the variables in the observed group and the interested group, respectively. Confounder variables affect units' treatment choices, which leads to the selection bias. 
In the medicine example, age is a confounder variable, so that people of different ages have different treatment preferences. Fig~\ref{fig: confounder} shows the age distribution of the observed treated/control group. Apparently, the age distribution of the observed treated group is different from the age distribution of the observed control group. This phenomenon exacerbates the difficulty of counterfactual outcome estimation as we need to estimate the control outcome of units in the treated group based on the observed control group, and similarly, estimate the treated outcome of units in the control group based on the observed treated group. If we directly train the potential outcome estimation model $\hat{Y}(x, w) =f_w(x)$ on the data with $W = w$ without handling the selection bias, the trained model would work poorly in estimating the potential outcome of $W = w$ for the units in the other group. This problem brought by the selection is also named as covariate shift in the Machine Learning community. 

Handing the problems caused by confounder variables is the essential part of causal inference, and the procedure of handing confounder variables is called \textit{adjust confounders}. The following part of this section briefly discusses the general solutions to tackle the above two problems caused by confounders under the ignorability assumption. The problem when there exists unobserved confounders will be discussed in Section~\ref{subsection: Unconfoundedness assumption}.

To solve the spurious effect problem, we should take the effect of confounder variables on outcomes into consideration. 
A general approach along this direction first estimates the treatment effect conditioning on the confounder variables and then conducts weighted averaging over the confounder according to its distribution. To be more specific,  
\begin{equation}
\begin{array}{rl}
    \hat{\text{ATE}} 
    &= \sum_{x}p(x)\mathbb{E}[Y^F|X = x, W = 1] - \sum_{x}p(x)\mathbb{E}[Y^F|X = x, W = 0] \\
    &= \sum_{\mathcal{X}^*}p(X \in \mathcal{X}^*)\left(\frac{1}{N_{\{i: X_i \in \mathcal{X}^*, W_i = 1\}}} \sum_{\{i: X_i \in \mathcal{X}^*, W_i = 1\}} Y_i^F\right) - \sum_{\mathcal{X}^*}p(X \in \mathcal{X}^*)\left(\frac{1}{N_{\{j: X_j \in \mathcal{X}^*, W_j = 1\}}} \sum_{\{j: X_j \in \mathcal{X}^*, W_j = 0\}} Y_j^F\right),
\end{array}
\label{eqn: adjust confounding_stratification}
\end{equation}
where $\mathcal{X}^*$ is a set of $X$ values, $p(X \in \mathcal{X}^*)$ is the probability of the background variables in $\mathcal{X}^*$ over the whole population, $\{i: x_i \in \mathcal{X}^*, W_i = w\}$ is the subgroup of units whose background variable values belong to $\mathcal{X}^*$ and treatment is equal to $w$. Stratification, which will be discussed in details later, is a representative method of this category.

For the selection bias problem, there are two general approaches to solve it. The first general approach handles selection bias by creating a pseudo group which is approximately close to the interested group. Possible methods include sample re-weighting, matching, tree-based methods, confounder balancing, balanced representation learning methods, multi-task based methods. The created pseudo-group alleviates the negative influence of the selection bias, and better counterfactual outcome estimations can be obtained. The other general approach first trains the base potential outcome estimation models solely on the observed data, and then correct the estimation bias caused by the selection bias. Meta-learning based methods belong to this category. 

\section{Causal Inference Methods Relying on Three assumptions}
\label{Section-3}

In this section, we introduce existing causal inference methods that rely on the three assumptions introduced in Section~\ref{sec2}. According to the way to control confounders, we divide these methods into the following categories: (1) Re-weighting methods; (2) Stratification methods; (3) Matching methods; (4) Tree-based methods; (5) Representation based methods; (6) Multi-task methods; and (7) Meta-learning methods.

\subsection{Re-weighting Methods}
\label{subsection: ample Re-weighting}
Due to the existence of confounders, the covariate distributions of the treated group and control group are different, which leads to the \textit{selection bias} problem as described in Section~\ref{genSolution}. In other words, the treatment assignment is correlated with covariates in the observational data. Sample re-weighting is an effective approach to overcome the selection bias. By assigning appropriate weight to each unit in the observational data, a pseudo-population can be created on which the distributions of the treated group and control group are similar. 

In sample re-weighting methods, a key concept is \textit{balancing score}. Balancing score $b(x)$ is a general weighting score, which is the function of $x$ satisfying: $W \independent x | b(x)$~\cite{imbens2015causal_book}, where $W$ is the treatment assignment and $x$ is the background variables. There are various designs of the balancing score, and apparently, the most trivial balancing score is $b(x) = x$.
Besides, propensity score is also a special case of balancing score. 
\begin{definition}
\textit{Propensity score: } The propensity score is defined as the conditional probability of treatment given background variables~\cite{rosenbaum1983central}:
\begin{equation}
\begin{array}{rl}
    e(x)=Pr(W=1|X=x)
\end{array}
\label{Eqn: propensity score}
\end{equation}
\end{definition}
In detail, a propensity score indicates the probability of a unit being assigned to a particular treatment given a set of observed covariates. Balancing scores that incorporate propensity score are the most common approach.

\begin{figure}
    \centering
    \includegraphics[width=0.95\textwidth]{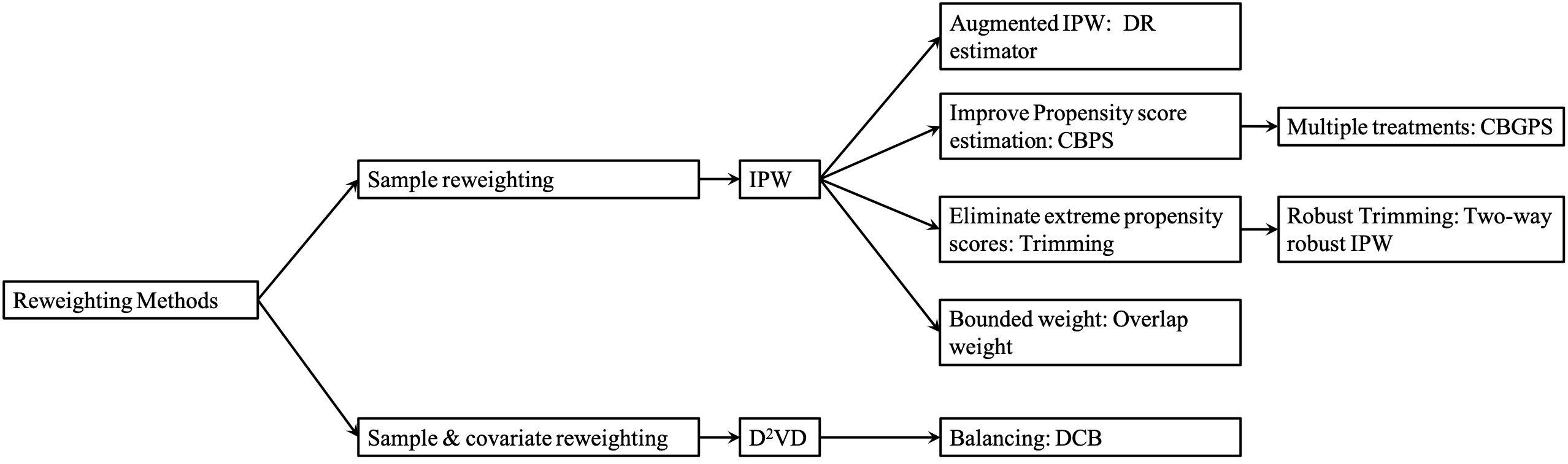}
    \caption{Categorization of Reweighting Methods}
    \label{fig:reweighting}
\end{figure}

A summarization of the algorithms mentioned in this section is shown in Fig~\ref{fig:reweighting}. The propensity score based sample reweighting will be introduced in the next section, followed by methods that weigh both samples and the covariates.

\subsubsection{Propensity score based sample re-weighting}
\label{subsection: sample re-weighting}

Propensity scores can be used to reduce selection bias by equating groups based on these covariates. Inverse propensity weighting (IPW)~\cite{rosenbaum1987model_IPW, rosenbaum1983central}, also named as inverse probability of treatment weighting (IPTW), assigns a weight $r$ to each sample:
\begin{equation}
\begin{array}{r}
    r =\frac{W}{e(x)}+\frac{1-W}{1-e(x)},
\end{array}
\end{equation}
where $W$ is the treatment assignment ($W = 1$ denotes being treated group; $W = 0$ denotes the control group), and $e(x)$ is the propensity score defined in Eqn.~\eqref{Eqn: propensity score}. 

After re-weighting, the IPW estimator of average treatment effect (ATE) is:
\begin{equation}
    \hat{\text{ATE}}_{IPW} = \frac{1}{n}\sum_{i = 1}^{n} \frac{W_i Y_i^F}{\hat{e}(x_i)} - \frac{1}{n}\sum_{i = 1}^{n} \frac{(1-W_i) Y_i^F}{1-\hat{e}(x_i)},
    \label{Eqn: IPW estimator}
\end{equation}
and its normalized version, which is preferred especially when the propensity scores are obtained by estimation~\cite{imbens2004nonparametric_review}:
\begin{equation}
    \hat{\text{ATE}}_{IPW} = \sum_{i = 1}^{n} \frac{W_i Y_i^F}{\hat{e}(x_i)}\bigg/\sum_{i = 1}^{n} \frac{W_i }{\hat{e}(x_i)}  - \sum_{i = 1}^{n} \frac{(1-W_i) Y_i^F}{1-\hat{e}(x_i)}\bigg/\frac{(1-W_i) }{1-\hat{e}(x_i)}.
    \label{Eqn: IPW normalized estimator}
\end{equation}
Both large and small sample theory show that adjustment for the scalar propensity score is enough to remove bias due to all observed covariates ~\cite{rosenbaum1983central}. The propensity score can be used to balance the covariates in the treatment and control groups and therefore reduce the bias through matching, stratification (subclassification), regression adjustment, or some combination of all three. \cite{d1998propensity} discusses the use of propensity score to reduce the bias, which also provides examples and detailed discussions.

However, in practice, the correctness of the IPW estimator highly relies on the correctness of the propensity score estimation, and slightly misspecification of propensity scores would cause ATE estimation error dramatically~\cite{imai2014covariate_balancing_ps}. To handle this dilemma, Doubly Robust estimator (DR)~\cite{robins1994estimation}, also named as Augmented IPW (AIPW), is proposed. DR estimator combines the propensity score weighting with the outcome regression, so that the estimator is robust even when one of the propensity score or outcome regression is incorrect (but not both). In detail, the DR estimator is formalized as:
\begin{equation}
\begin{split}
    \hat{\text{ATE}}_{DR} &= \frac{1}{n}\sum_{i = 1}^{n}\left\{ \left[\frac{W_i Y_i^F}{\hat{e}(x_i)} -\frac{W_i -\hat{e}(x_i) }{\hat{e}(x_i)} \hat{m}(1, x_i)\right] - \left[\frac{(1-W_i) Y_i^F}{1-\hat{e}(x_i)} -\frac{W_i -\hat{e}(x_i) }{1-\hat{e}(x_i)} \hat{m}(0, x_i)\right] \right\}\\
    &= \frac{1}{n} \sum_{i = 1}^{n} \left\{\hat{m}(1, x_i) + \frac{W_i(Y_i^F - \hat{m}(1, x_i))}{\hat{e}(x_i)} - \hat{m}(0, x_i) - \frac{(1-W_i)(Y_i^F - \hat{m}(0, x_i))}{1-\hat{e}(x_i)}
    \right\},
\end{split}
\end{equation}
where $\hat{m}(1, x_i)$ and $\hat{m}(0, x_i)$ are the regression model estimations of treated and control outcomes. The DR estimator is consistent and therefore asymptotically unbiased, if either the propensity score is correct or the model correctly reflects the true relationship among exposure and confounders with the outcome~\cite{fan2016improving}. In reality, one definitely cannot guarantee whether one model can accurately explain the relationship among variables. The combination of outcome regression with weighting by propensity score ensures that the estimators are robust to misspecification of one of these models ~\cite{bang2005doubly,robins2007comment,robins1994estimation,scharfstein1999comments}.

The DR estimator consults outcomes to make the IPW estimator robust when propensity score estimation is not correct. An alternative way is to improve the estimation of propensity scores. In the IPW estimator, propensity score serves as both the probability of being treated and the covariate balancing score, 
covariate balancing propensity score (CBPS)~\cite{imai2014covariate_balancing_ps} is proposed to exploit such dual characteristics. In particular, CBPS estimates propensity scores by solving the following problem:
\begin{equation}
    \mathbb{E}\left[ \frac{W_i\Tilde{x_i}}{e(x_i; \beta)} - \frac{(1-W_i)\Tilde{x_i}}{1-e(x_i; \beta)} \right] = 0,
    \label{eqn: CBPS}
\end{equation}
where $\Tilde{x_i} = f(x_i)$ is a predefined vector-valued measurable function of $x_i$. By solving the above problem, CBPS 
directly constructs the covariate balancing score from the estimated parametric propensity score, which increase the robustness to the misspecification of the propensity score model. An extension of CBPS is the covariate balancing generalized propensity score (CBGPS)~\cite{fong2018covariate_CBGPS}, which enables to handle the treatment with continuous value. Due to the continuous valued treatment, it's hard to directly minimized the covariates distribution distance between the control and treated group. CBGPS solves this problem by mitigating the definition of the balancing score. Based on the definition, the treatment assignment is conditionally independent of the background variables, CBGPS directly minimize the correlation between the treatment assignment and the covariates after weighting. In specific, the objective of CBGPS is to learn a propensity score based weight so that the weighted correlation between the treatment assignment and the covariates are minimized:
\begin{equation}
\begin{split}
    \mathbb{E}\left( \frac{p(t^*)}{p(t^*|x^*)} t^* x^*\right) &= 
    \int\left\{\int\ \frac{p(t^*)}{p(t^*|x^*)}t^* dP(t^*|x^*)\right\}x^*dP(x^*)\\
    &=\mathbb{E}(t^*)\mathbb{E}(x^*) = 0,\\
\end{split}
\label{eqn: CBGPS}
\end{equation}
where $p(t^*|x^*)$ is the propensity score, and $\frac{p(t^*)}{p(t^*|x^*)}$ is the balancing weight, $t^*$ and $x^*$ is the treatment assignment and the background variables after centering and orthogonalizing (i.e., normalization). In a nutshell, both CBPS and CBGPS learns the propensity score based sample weight directly towards the covariate balancing goal, which can alleviates negative effect brought by model misspecification of propensity score.

Another drawback of the IPW estimator is that it might be unstable if the estimated propensity scores are small. If the probability of either treatment assignment is small, the logistic regression model can become unstable around the tails, causing the IPW to also be less stable. To overcome this issue, trimming is routinely employed as a regularization strategy, which eliminates the samples whose propensity scores are less than a pre-defined threshold~\cite{lee2011weight_trimming}. However, this approach is highly sensitive to the amount of trimming~\cite{ma2010robust_IPW}. Also, theoretical results in~\cite{ma2010robust_IPW} show that the small probability of propensity scores and the trimming procedure may result in different non-Gaussian asymptotic distribution of IPW estimator. Based on this observation, a two-way robustness IPW estimation algorithm is proposed in~\cite{ma2010robust_IPW}. This method combines subsampling with a local polynomial regression based trimming bias corrector, so that it is robust to both small propensity score and the large scale of trimming threshold. An alternative approach to overcome the instability of IPW under small propensity scores is to redesign the sample weight so that the weight is bounded. In~\cite{li2018balancing}, the overlap weight is proposed, in which each unit's weight is proportional to the probability of that unit being assigned to the opposite group. In detail, the overlap weight $h(x)$ is defined as $h(x) \propto 1-e(x)$, where $e(x)$ is the propensity score. The overlap weight is bounded within the interval $[0,0.5]$, and thus it is less sensitive to extreme vale of propensity score. Recent theoretical results show that the overlap weight has the minimum asymptotic variance among all balancing weights~\cite{li2018balancing}.

\subsubsection{Confounder balancing}
The aforementioned sample re-weighting methods could achieve balance in the sense that the observed variables are considered equally as confounders. However, in real cases, not all the observed variables are confounders. Some of the variables, named as adjustment variables, are only predictive to the outcome, and some others might be irrelevant variables. Adjusting on the adjustment variables by Lasso, although it cannot reduce the bias, helps decrease the variance~\cite{sauer2013review_covariate_selection,bloniarz2016lasso_adjustment}. While including the irrelevant variables would cause overfitting.

Based on the separateness assumption that the observed variables can be decomposed into confounders, adjusted variables and the irrelevant variables, in~\cite{kuang2017treatment_aaai}, the Data-Driven Variable
Decomposition (D$^2$VD) algorithm is proposed to distinguish the confounders and adjustment variables, and meanwhile, eliminate the irrelevant variables. 
In detail, the adjusted outcome is written as:
\begin{equation}
    Y^{*}_{\text{D}^2\text{VD}} = \left(Y^{F} - \phi(\mathbf{z})\right) \frac{W-p(x)}{p(x)(1-p(x))},
\end{equation}
where $\mathbf{z}$ is the adjustment variables. 
Therefore, the ATE estimator of $\text{D}^2\text{VD}$ is:
\begin{equation}
     \text{ATE}_{\text{D}^2\text{VD}} = \mathbb{E}\left[ \left(Y^{F} - \phi(\mathbf{z})\right) \frac{W-p(x)}{p(x)(1-p(x))}  \right].
\end{equation}
To get $\text{ATE}_{\text{D}^2\text{VD}}$, the 
$Y^{*}_{\text{D}^2\text{VD}}$ is regressed on all observed 
variables. The objective function is $l_2$ loss between $Y^{*}_{\text{D}^2\text{VD}}$ 
and the linear regression function on all observed variables, along with sparse regularization to distinguish the confounder, adjusted variables, and irrelevant variables. However, little prior knowledge about the interactions among observed variables is provided in practice, and the data are usually high-dimensional and noisy. To solve this, Differentiated Confounder Balancing (DCB) algorithm~\cite{kuang2017estimating_condounder_balancing_kdd} is proposed to select and differentiate confounders to balance the distributions. Overall, DCB balances the distributions by re-weighting both the samples and confounders.

\subsection{Stratification Methods}
Stratification, also named as \textit{subclassification} or \textit{blocking}~\cite{imbens2015causal_book}, is a representative method to adjust the confounders.
The idea of stratification is to adjust the bias that stems from the difference between the treated group and the control group by splitting the entire group into homogeneous subgroups (blocks). Ideally, in each subgroup, the treated group and the control group are similar under certain measurements over the covariates, therefore, the units in the same subgroup can be viewed as sampled from the data under randomized controlled trials. Based on the homogeneity of each subgroup, the treatment effect within each subgroup (i.e., CATE) can be calculated through the method developed on RCTs data. 
After having the CATE of each subgroup, the treatment effect over the interested group can be obtained by combining the CATEs of subgroups belonging to that group, as shown in~\eqref{eqn: adjust confounding_stratification}. In the following, we adopt the calculation of ATE as an example. 
In detail, if we separate the whole dataset into $J$ blocks, the ATE is estimated as:
\begin{equation}
    \text{ATE}_{\text{strat}} = \hat{\tau}^{\text{strat}} = \sum_{j = 1}^{J}q(j)\left[\bar{Y}_t(j) -\bar{Y}_c(j)\right],
\end{equation}
where $\bar{Y}_t(j)$ and $\bar{Y}_c(j)$ are the average of the treated outcome and control outcome in the $j$-th block, respectively. $q(j) = \frac{N(j)}{N}$ is the portion of the units in the $j$-th block to the whole units.

Stratification effectively decreases the bias of ATE estimation compared with the difference-estimator where ATE is estimated as: $\text{ATE}_{\text{diff}} = \hat{\tau}^{diff} = \frac{1}{N_t} \sum_{i: W_i = 1} Y_i^{CF} - \sum_{i: W_i = 0} Y_i^{CF}$. In particular, if we assume the outcome is linear with the covariates, i.e., $\mathbb{E}[Y_i(w)|X_i = x] = \alpha + \tau* w + \beta * x$. The bias of the difference-estimator is:  
\begin{equation}
    \mathbb{E}[\hat{\tau}^{\text{diff}} - \tau | X, W] = (\bar{X}_t - \bar{X}_c)\beta.
\end{equation}
While, the bias of the stratification estimator is the weighted average of the within-block bias:
\begin{equation}
    \mathbb{E}[\hat{\tau}^{\text{strat}} - \tau | X, W] = \left(\sum_{j = 1}^{J}q(j)\left(\bar{X}_t(j) - \bar{X}_c(j)\right)\right)\beta.
\end{equation}
Compared with the difference estimator, the stratification estimator reduces the bias per covariate by the factor:
\begin{equation}
    \gamma_k = \frac{\sum_j q(j) \left( \bar{X}_{t,k}(j) - \bar{X}_{c,k}(j)  \right)}{\bar{X}_{t,k} - \bar{X}_{c,k}},
\end{equation}
where $\bar{X}_{t,k}(j)$ ($\bar{X}_{c,k}(j)$) is the average of $k$-th covariate of treated (control) group in $j$-th block, and $\bar{X}_{t,k}$ ($\bar{X}_{c,k}$) is the average of $k$-th covariate in the whole treated (control) group.

The key component of stratification methods is how to create the blocks and how to combine the created blocks. The equal frequency~\cite{rosenbaum1983central} is a common strategy to create blocks. Equal-frequency approach split the block by the appearance probability, such as the propensity score, so that the covariates have the same appearance probability (i.e., the propensity score) in each subgroup (block). The ATE is estimated by weighted average of each block's CATE, with the weight as the fraction of the units in this block. 
However, this approach suffers from high variance due to the insufficient overlap between treated and control groups in the blocks whose propensity score is very high or low. To reduce the variance, in~\cite{hullsiek2002propensity_strat}, the blocks, which divided according to the propensity score, are re-weighted by the inverse variance of the block-specific treatment effect. Although this method reduces the variance of equal-frequency method, it unavoidably increases the estimation bias. 

The stratification methods described above are all splitting the blocks according to the pre-treatment variables. However, in some real-world applications, it is required to compare the outcome conditioned on some post-treatment variables, denoted as $S$. For example, the ``surrogate'' markers of disease progression (i.e., intermediate outcome) like CD4 count and measures of viral load in AIDS are the post-treatment variables~\cite{frangakis2002principal_stratification}. In the studies comparing drugs for AIDS patients, the researchers are interested in the effect of AIDS drugs on group with CD4 count lower than $200$ cell/$\text{mm}^3$. However, directly comparing the observed outcomes on the group with $S^{obs}<200$ is not the true effect because the compared two subgroups: $\{i: W_i = 1,  S^{obs}<200\}$ and $\{j: W_j = 0, S^{obs}<20\}$, where $S^{obs}$ is the observed post-treatment values, have great discrepency if the treatment has effect on the intermediate results. To solve this, principle stratification~\cite{frangakis2002principal_stratification} constructs the subgroup based on the potential values of the pre-treatment variables. Analogous to the potential outcome defined in~\ref{subsec:definition}, potential pre-treatment variables value, denoted as $S(W = w)$, is the potential value of $S$ under treatment with value $w$. With the nature assumption that potential value of $S$ is independent of the treatment assignment, the treatment effect of subgroup can be obtained by comparing the outcomes of two sets: $\{Y_i^{obs}: W_i = 1, S_i(W_i = 1) = v_1,  S_i(W_i = 0) = v_2\}$ and $\{Y_j^{obs}: W_j = 0, S_j(W_j = 1) = v_1, S_j(W_j = 0) = v_2\}$, where $v_1$ and $v_2$ are two post-treatment values. The comparison based on the potential values of post-treatment variables ensures that the compared two set are similar, so that the obtained treatment effect is the true effect. 

\subsection{Matching Methods}
As mentioned previously, missing counterfactuals and confounder bias are two major challenges in treatment effect estimation. Matching based approaches provide a way to estimate the counterfactual and, at the same time, reduce the estimation bias brought by the confounders. In general, the potential outcomes of the $i$-th unit estimated by matching are~\cite{abadie2004implementing_matching_stata}: 
\begin{equation}
\label{Eqn: Matching estimator}
    \begin{array}{cc}
        \hat{Y}_{i}(0)  = \left\{\begin{array}{lr}
            Y_i & \text{if }W_i = 0,  \\
            \frac{1}{\# \mathcal{J}(i)}\sum_{l\in\mathcal{J}(i)} Y_l & \text{if }W_i = 1 ;
        \end{array} \right. &  
         \hat{Y}_{i}(1)  = \left\{\begin{array}{lr}
            \frac{1}{\# \mathcal{J}(i)}\sum_{l\in\mathcal{J}(i)} Y_l & \text{if }W_i = 0, \\
            Y_i & \text{if }W_i = 1;  \\
        \end{array} \right.
        \\
    \end{array}
\end{equation}
where $\hat{Y}_{i}(0)$ and $\hat{Y}_{i}(1)$ are the estimated control and treated outcome, $\mathcal{J}(i)$ is the matched neighbors of unit $i$ in the opposite treatment group~\cite{austin2011introduction}.

The analysis of the matched sample can mimic that of an RCT: one can directly compare outcomes between the treated and control group within the matched sample. In the context of an RCT, one expects that, on average, the distribution of covariates will be similar between treated and control groups. Therefore, matching can be used to reduce or eliminate the effects of confounding when using observational data to estimate treatment effects~\cite{austin2011introduction}. 

\subsubsection{Distance Metric}
Various distances have been adopted to compare the closeness between units~\cite{gu1993comparison}, such as the widely used Euclidean distance~\cite{rubin1973matching_NNM} and Mahalanobis distance~\cite{rubin2000combining}. Meanwhile, many matching methods develop their own distance metrics, which can be abstracted as: $D(\mathbf{x}_i, \mathbf{x}_j) = ||f(\mathbf{x}_i) - f(\mathbf{x}_j)||_2$. The existing distance metrics mainly vary in how to design the transformation function $f(\cdot)$.

\textit{Propensity score based transformation.}
Original covariates of units can be represented by propensity scores. As a result, the similarity between two units can be directly calculated as: $D(\mathbf{x}_i, \mathbf{x}_j) = |e_i - e_j|$, where $e_i$, and $ e_j$ are the propensity scores of $\mathbf{x}_i$ and $\mathbf{x}_j$, respectively. Later, the linear propensity score based distance metric is also proposed, which is defined as $D(\mathbf{x}_i, \mathbf{x}_j) = |\text{logit}(e_i) - \text{logit}(e_j)|$. This improved version is recommended since it can effectively reduce the bias~\cite{stuart2010matching_review}. Furthermore, the propensity score based distance metric can be combined with other existing distance metrics, which provides a fine-grained comparison. In~\cite{rubin2000combining}, when the difference of two unit's propensity scores is within a certain range, they are further compared with other distances on some key covariates. Under this metric, the closeness of two units contains two criteria: they are relatively close under propensity score measure, and they particularly similar under the comparison of the key covariates~\cite{stuart2010matching_review}.

\textit{Other transformations.} Propensity score only adopts the covariate information, while some other distance metrics are learned by utilizing both the covariates and the outcome information so that the transformed space can preserve more information. One representative metric is the prognosis score~\cite{hansen2008prognostic}, which is the estimated control outcome. The transformation function is represented as: $f(x) = \hat{Y}_c$. However, the performance of the prognosis score relies on modeling the relationship between the covariates and control outcomes. Moreover, the prognosis score only takes the control outcome into consideration and ignores the treated outcome. The Hilbert-Schmidt Independence Criterion based nearest neighbor matching (HSIC-NNM) proposed in~\cite{chang2017informative_subspace_aaai17} could overcome the drawbacks of prognosis score. HSIC-NNM learns two linear projections for control outcome estimation task and treated outcome estimation task separately. To fully explore the observed control/treated outcome information, the parameters of linear projection is learned by maximizing the nonlinear dependency between the projected subspace and the outcome: $M_w = \argmax_{M_w}\text{HSIC}(\mathbf{X}_{w} M_w, Y_w^{F}) - \mathcal{R}(M_w)$, where $w = 0, 1$ represent the control group and treated group, respectively. $\mathbf{X}_{w} M_w$ is the transformed subspace with the transformation function as: $f(x) = x M_w $. $Y_w^{F}$ is the observed control/treated outcome, and $\mathcal{R}$ is the regularization to avoid overfitting. The objective function ensures the learned transformation functions project the original covariates to an information subspace where similar units will have similar outcomes. 

Compared with propensity score based distance metric that focuses on balancing, prognosis score and HSIC-NNM focus on embedding the relationship between the transformed space and the observed outcome. These two lines of methods have different advantages, and some recent work tries to integrate these advantages together. 
In~\cite{li2017matching_nips17}, the balanced and
nonlinear representation (BNR) is proposed to project the covariates into a balanced low-dimensional space. In detail, the parameters in the nonlinear transformation function is learned by jointly optimizing the following two objectives: (1) Maximizing the differences of noncontiguous-class scatter and within-class scatter so that the units with the same outcome prediction shall have similar representations after transformation; (2) Minimizing the maximum mean discrepancy between the transformed control and outcome group in order to get the balanced space after transformation. A series of works that have similar objectives but vary in balancing regularization have been proposed, such as using the conditional generative adversarial network to ensure the transformation function blocks the treatment assignment information~\cite{yao2019estimation,lee2018ITE_adversarial}.

The methods mentioned above adopt either one or two transformations for treated and control groups separately. Different from the existing method, Randomized Nearest Neighbor
Matching (RNNM)~\cite{li2016matching_digital_ijcai16} adopts a number of random linear projections as the transformation function, and the treatment effects are obtained as the median treatment effect by nearest-neighbor matching in each transformed subspace. The theoretical motivation of this approach is the Johnson-Lindenstrauss (JL) lemma, which guarantees that the pairwise similarity information of the points in the high-dimensional space can be preserved through random linear projection. Powered by the JL lemma, RNNM ensembles the treatment effect estimation results of several linear random transformations.

\subsubsection{Choosing a Matching Algorithm}
\begin{figure}
\includegraphics[width=0.95\textwidth]{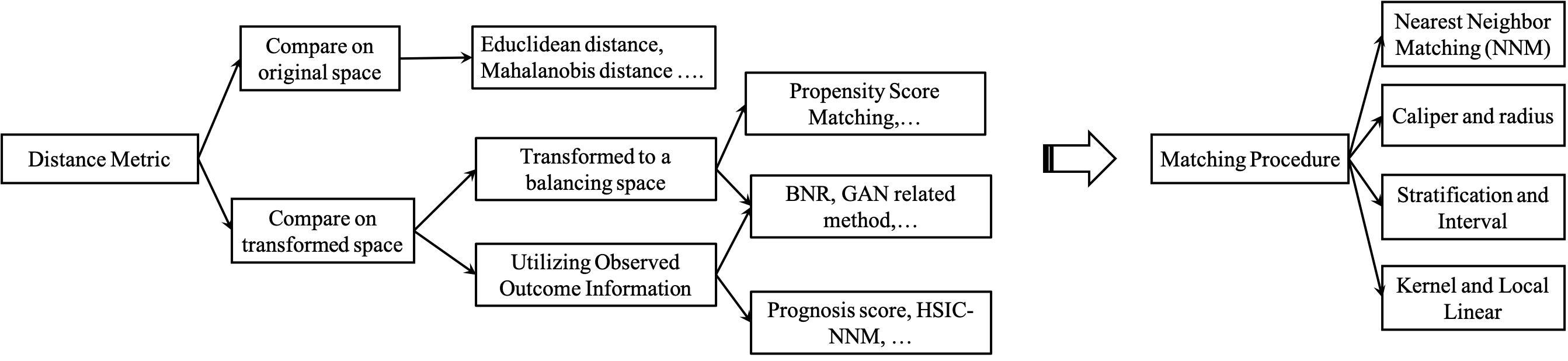}
\centering
\caption{Categorization of Matching Methods.}
\label{fig: matching}
\end{figure}

After defining the similarity metric, the next step is to find the neighbors. In~\cite{caliendo2008some}, existing matching algorithms are divided into four essential approaches, including the nearest neighbor matching, caliper, stratification and kernel, as shown in Fig.~\ref{fig: matching}. The most straightforward matching estimator is nearest neighbour matching (NNM). In particular, a unit in the control group is chosen as the matching partner for a treated unit, so that they are closest based on a similarity score (e.g., propensity score). The NNM has several variants like NNM with replacement and NNM without replacement. Treated units are matched to one control, called pair matching or 1-1 matching, or treated units are matched to two controls, called 1-2 matching, and so on. It's a trade-off to determine the number of neighbors, since a large number of neighbors may result in the treatment effect estimator with high bias but low variance, while small number results in low bias but high variance. It is known, however, that the optimal structure is a full matching in which a treated unit may have one or several controls or a control may have one or several treated units~\cite{gu1993comparison}. 

NNM may have bad matches if the closest partner is far away. One can set a tolerance level on the maximum propensity score distance (caliper) to avoid this problem. Hence, caliper matching is one form of imposing a common support condition. 

The stratification matching is to partition the common support of the propensity score into a set of intervals and then to take the mean difference in outcomes between treated and control observations in order to calculate the impact within each interval. This method is also known as interval matching, blocking and subclassification~\cite{rosenbaum1985constructing}. 

The matching algorithms discussed above have in common that only a few observations in the control group are used to create the counterfactual outcome of a treatment observation. Kernel matching (KM) and local linear matching (LLM) are nonparametric matching that use weighted averages of observations in the control group to create the counterfactual outcome. Thus, one major advantage of these approaches is the lower variance, because we use more information to create counterfactual outcome. 

Here, we also want to introduce another matching method called Coarsened Exact Matching (CEM) proposed in~\cite{iacus2012causal_CEM}. Because either the 1-k matching or the full matching fails to consider the extrapolation region, where few or no reasonable matches exist in the other treatment group, CEM was proposed to handle this problem. CEM first coarsen the selected important covariate,i.e., discretization, and then perform exact matching on the coarsened covariates. For example, if the selected covariates are age ($\text{age}>50$ is $1$, and others are $0$) and gender (female as $1$, and male as $0$). A female patient with age $50$ in the treated group is represented by the coarsen covariates as $(1, 1)$. She will only match the patients in the treated group with exactly the same coarsened covariates value. After exact matching, the whole data is separated into two subsets. In one subset, every unit has its exact matched neighbors and it is the opposite in the other subset which contains the units in the extrapolation region. The outcomes of units in the extrapolation region are estimated by the outcome prediction model trained on the matched subset. So far, the treatment effect on the two subsets can be estimated separately, and the final step is to combine treatment effect on the two subsets by weighted average. 

We have provided several different matching algorithms, but the most important question is how we should select a perfect matching method. Asymptotically all matching methods should yield the same results as the sample size grows and they will become closer to comparing only exact matches ~\cite{smith2000critical}. When we only have small samples size, this choice will be important ~\cite{heckman1998matching_matching_variable}. There is one trade-off between bias and variance.

\subsubsection{Variables to include}
The above two subsections illustrate the key steps in matching procedure, and in this subsection, we briefly discuss what kinds of variables should be included in the matching, a.k.a feature selection, to improve the matching performance.
Many literatures~\cite{rubin1996matching_matching_variable, glazerman2003nonexperimental_matching_variable, heckman1998matching_matching_variable} suggest to include as many variables that are related to the treatment assignment and the outcome as possible, in order to satisfy the strong ignorablity assumption. However, post-treatment variables, which are the variables affected by the treatment assignment, should be excluded in the matching procedure~\cite{rosenbaum1984reducing_subclassification}. Moreover, besides the post-treatment variables, researchers also suggest excluding the instrumental variables~\cite{pearl2012bias-amplifying,wooldridge2016should}, because they tend to amplify the bias of treatment effect estimator.  

\subsection{Tree-based Methods}
Another popular method in causal inference is based on decision tree learning, which is one of the predictive modeling approaches. Decision tree is a non-parametric supervised learning method used for classification and regression. The goal is to create a model that predicts the value of a target variable by learning simple decision rules inferred from data. 

Tree models where the target variable is discrete are called classification trees with prediction error measured based on misclassification cost. In these tree structures, leaves represent class labels and branches represent conjunctions of features that lead to those class labels. Decision trees where the target variable is continuous are called regression trees with prediction error measured by the squared difference between the observed and predicted values. The term Classification And Regression Tree (CART) analysis is an umbrella term used to refer to both of the above procedures~\cite{breiman2017classification}. In CART model, the data space is partitioned and a simple prediction model for each partition space is fitted, and therefore every partitioning can be represented graphically as a decision tree ~\cite{loh2011classification}. 
 
For estimating heterogeneity in causal effects, a data-driven approach ~\cite{athey2016recursive_hce} based on CART is provided to partition the data into subpopulations that differ in the magnitude of their treatment effects. The valid confidence intervals can be created for treatment effects, even with many covariates relative to the sample size, and without "sparsity" assumptions. This approach is different from conventional CART in two aspects. First, it focuses on estimating conditional average treatment effects instead of directly predicting outcomes as in the conventional CART. Second, different samples are used for constructing the partition and estimating effects each subpopulation, which is referred to as the honest estimation. However, in conventional CART, the same samples are used for these two tasks. 

In CART, a tree is built up until a splitting tolerance is reached. There is only one tree, and it is grown and pruned as needed. However, BART is an ensemble of trees, so it is more comparable to random forests. A Bayesian ``sum-of-trees'' model called Bayesian Additive Regression Trees (BART) is developed in~\cite{chipman2007bayesian} ~\cite{chipman2010bart}. Every tree in BART model is a weak learner, and it is constrained by a regularization prior. Information can be extracted from the posterior by a Bayesian backfitting MCMC algorithm. BART is a nonparametric Bayesian regression model, which uses dimensionally adaptive random basis elements. Let $W$ be a binary tree which has a set of interior node decision rules and terminal nodes, and let $M = \{\mu_1 , \mu_2, ..., \mu_B\}$ be parameters associated with each of the $B$ terminal nodes for $W$. We use $g(x; W, M)$ to assign a $\mu_b \in M$ to input vector $x$. The sum-of-trees model can be expressed as: 
\begin{equation}
    Y=g(x;W_1,M_1) + g(x;W_2,M_2) + \cdots + g(x;W_m,M_m) + \varepsilon,
\end{equation}
\begin{equation}
    \varepsilon \sim N(0, \sigma^{2}),
\end{equation}

BART has a couple of advantages. It is very easy to implement and only needs to plug in the outcome, treatment assignment, and confounding covariates. In addition, it doesn't require any information about how these variables are parametrically related, so that it requires less guess when fitting the model. Moreover, it can deal with a mass of predictors, yield coherent uncertainty intervals, and handle continuous treatment variables and missing data~\cite{hill2011bayesian}. 

BART is proposed to estimate average causal effects. In fact, it can also be used to estimate individual-level causal effects. BART not only can easily identify the heterogeneous treatment effects, but also get more accurate estimates of average treatment effects compared to other methods like propensity score matching, propensity score weighting, and regression adjustment in the nonlinear simulation situations examined ~\cite{hill2011bayesian}. 

In most previous methods, the prior distribution over treatment effects is always induced indirectly, which is difficult to be attained. A flexible sum of regression trees (i.e., a forest) can address this issue by modeling a response variable as a function of a binary treatment indicator and a vector of control variables~\cite{hahn2017bayesian}. This approach interpolates between two extremes: entirely and separately modeling the conditional means of treatment and control, or only the treating treatment assignment as another covariate. 

Random forest is a classifier consisting of a combination of tree predictors, in which each tree depends on a random vector that is independently sampled and has the identical distribution for all trees~\cite{breiman2001random}. This model can also be extended to estimate heterogeneous treatment effects based on the Breiman's random forest algorithm ~\cite{causal_forest}. Trees and forests can be considered as nearest neighbor methods with an adaptive neighborhood metric. Tree-based methods seek to find training examples that are close to a point $x$, but now closeness is defined with respect to a decision tree. And the closest points to $x$ are those that fall in the same leaf as it. The advantage of using trees is that their leaves can be narrower along the directions where the signal is changing fast and wider along the other directions, potentially leading to a substantial increase in power when the dimension of the feature space is even moderately large.

The tree-based framework also can be extended to uni- or multi-dimensional treatments~\cite{wang2015robust_complex_ad}. Each dimension can be discrete or continuous. A tree structure is used to specify the relationship between user characteristics and the corresponding treatment. This tree-based framework is robust to model misspecification and highly flexible with minimal manual tuning.

\subsection{Representation Learning Methods}
\subsubsection{Balanced representation learning}

The most basic assumption used in statistical learning theory is that training data and test data are drawn from the same distribution. However, in most practical cases, the test data are drawn from a distribution that is only related, but not identical, to the distribution of the training data. In causal inference, this is also a big challenge. Unlike the randomized control trials, the mechanism of treatment assignment is not explicit in observational data. Therefore, interventions of interest are not independent of the property of the subjects. For example, in an observational study of the treatment effect of a medicine, the medicine is assigned to individuals based on several factors, including the known confounders and some unknown confounders. As a result, the counterfactual distribution will generally be different from the factual distribution. Thus, it is necessary to predict counterfactual outcomes by learning from the factual data, which converts the causal inference problem to a domain adaptation problem.

Extracting effective feature representations is critical for domain adaptation. A model ~\cite{ben2007analysis} with a generalization bound is proposed to formalize this intuition theoretically, which can not only explicitly minimize the difference between the source and target domains, but also maximize the margin of the training set. Building on this work ~\cite{ben2007analysis}, the discrepancy distance between distributions is tailored to adaptation problems with arbitrary loss functions~\cite{mansour2009domain}. In the following discussions, the discrepancy distance plays an important role in addressing the domain adaptation problem in causal inference.

So far, we can see a clear connection between counterfactual inference and domain adaptation. An intuitive idea is to enforce the similarity between the distributions of different treatment groups in the representation space. The learned representations trade-off three objectives: (1)  low-error prediction over the factual representation, (2) low-error prediction over counterfactual outcomes by taking into account relevant factual outcomes, and (3) the distance between the distribution of treatment population and that of control population~\cite{johansson2016learning}. 
Following this motivation,~\cite{shalit2017estimating} gives a simple and intuitive generalization-error bound. It shows that the expected ITE estimation error of representation is bounded by a sum of the standard generalization-error of that representation and the distance between the treated and control distributions based on representation. Integral probability metric (IPM) is used to measure the distances between distributions, and explicit bounds are derived for the Wasserstein distance and Maximum Mean Discrepancy (MMD) distance. The goal is to find a representation $\Phi: X \to R$ and hypothesis $h : X \times \{0, 1\} \to Y$ that
minimizes the following objective function:
\begin{equation}
   \min_{h,\Phi} \frac{1}{n}\sum_{i=1}^{n}r_i\cdot L(h(\Phi(x_i),W_i),y_i)+\lambda \cdot R(h) + \alpha \cdot IPM_G(\{\Phi(x_i)\})_{i:W_i=0},\{\Phi(x_i)\})_{i:W_i=1}),\\
\end{equation}
where $w_i=\frac{W_i}{2u}+\frac{1-W_i}{2(1-u)}$, $u=\frac{1}{n}\sum_{i=1}^{n}W_i$, and the weights $r_i$ compensate for the difference in treatment group size. $R$ is a model complexity term.  Given two probability density functions $p$, $q$ defined over $S \subseteq R^d$, and a function family $G$ of functions $g : S \to R$, the IPM is defined as:
\begin{equation}
    IPM_G(p,q):=\sup_{g\in G} |\int_S g(s)(p(s)-q(s))ds|.
\end{equation}

This model allows for learning complex nonlinear representations and hypotheses with large flexibility. When the dimension of $\Phi$ is high, it risks losing the influence of $t$ on $h$ if the concatenation of $\Phi$  and $W$ is treated as input. To address this problem, one approach is to parameterize $h_1(\Phi)$ and $h_0(\Phi)$ as two separate ``heads'' of the joint network. $h_1(\Phi)$ is used to estimate the outcome under treatment and $h_0(\Phi)$ is for the control group. Each sample is used to update only the head corresponding to the observed treatment. The advantage is that statistical power is shared in the common representation layers and the influence of treatment is retained in the separate heads ~\cite{shalit2017estimating}. This model can also be extended to any number of treatments, as described in the perfect match (PM) approach~\cite{schwab2018perfect}. Following this idea, a few improved models have been proposed and discussed. For example, ~\cite{johansson2018weighted_rep} brings together shift-invariant representation learning and re-weighting methods. \cite{hassanpour2019counterfactual_importance_ijcai19} presents a new context-aware weighting scheme based on the importance sampling technique, on top of representation learning, to alleviate the selection bias problem in ITE estimation. 

Existing ITE estimation methods mainly focus on balancing the distributions of control and treated groups, but ignore the local similarity information that provides meaningful constraints on the ITE estimation. In ~\cite{yao2018representation, yao19icdm}, a local similarity preserved individual treatment effect (SITE) estimation method is proposed based on deep representation learning. SITE preserves local similarity and balances data distributions simultaneously. The framework of SITE contains five major components: representation network, triplet pairs selection, position-dependent deep metric (PDDM), middle point distance minimization (MPDM), and the outcome prediction network. To improve the model efficiency, SITE takes input units in a mini-batch fashion, and triplet pairs could be selected from every mini-batch. The representation network learns latent embeddings for the input units. With the selected triplet pairs, PDDM and MPDM can preserve the local similarity information and meanwhile achieve the balanced distributions in the latent space. Finally, the embeddings of mini-batch are fed forward to a dichotomous outcome prediction network to get the potential outcomes. The loss function of SITE is as follows:
\begin{equation}
    L=L_{FL}+ \beta L_{PDDM} +\gamma L_{MPDM}+ \lambda||M||_2
\end{equation}
where $L_{FL}$ is the factual loss between the estimated and observed factual outcomes. $L_{PDDM}$ and $L_{MPDM}$ are the loss functions for PDDM and MPDM, respectively. The last term is $L_2$ regularization on model parameters $M$ (except the bias term).

Most models focus on covariates with numerical values, while how to handle covariates with textual information for treatment effect estimation is still an open question. One major challenge is how to filter out the nearly instrumental variables which are the variables more predictive to the treatment than the outcome. Conditioning on those variables to estimate the treatment effect would amplify the estimation bias. To address this challenge, a conditional treatment-adversarial learning based matching (CTAM) method is proposed in  ~\cite{yao2019estimation}. CTAM incorporates the treatment-adversarial learning to filter out the information related to nearly instrumental variables when learning the representations, and then it performs matching among the learned representations to estimate the treatment effects. The CTAM contains three major components: text processing, representation learning, and conditional treatment discriminator. Through the text processing component, the original text is transformed into vectorized representation $S$. After that, $S$ is concatenated with the non-textual covariates $X$ to construct a unified feature vector, which is then fed into the representation neural network to get the latent representation $Z$. After learning the representation, $Z$, together with potential outcomes $Y$, are fed into the conditional treatment discriminator. During the training procedures, the representation learner plays a minimax game with the conditional treatment discriminator: By preventing the discriminator from assigning correct treatment, the representation learner can filter out the information related to nearly instrumental variables. The final matching procedure is performed in the representation space $Z$. The conditional treatment-adversarial learning helps reduce the bias of treatment effect estimation.

Compared to the above regression-based methods after representation learning, matching method is more interpretable, because any sample's counterfactual outcome is directly set to be the factual outcome of its nearest neighbor in the group receiving the opposite treatment. Nearest neighbor matching (NNM) sets the counterfactual outcome of any treatment (control) sample to be equal to the factual outcome of its nearest neighbor in the control (treatment) group. Although being simple, flexible and interpretable, most NNM approaches could be easily misled by variables that do not affect the outcome. To address this challenge, matching can be performed on subspaces that are predictive of the outcome variable for both the treatment group and the control group. Applying NNM in the learned subspaces leads to a more accurate estimation of the counterfactual outcomes and therefore the accurate estimation of treatment effects. \cite{chang2017informative_subspace_aaai17} estimates the counterfactual outcomes of treatment samples by learning a projection matrix that maximizes the nonlinear dependence between the subspace and outcome variable for control samples. Then it directly applies the learned projection matrix to all the samples and finds every treatment sample's matched control sample in the subspace.

\subsection{Multitask Learning Methods}

Treatment group and control group always share some common features except for their idiosyncratic characteristics. Naturally, causal inference can be conceptualized as a multitask learning problem with a set of shared layers for treated group and control group together, and a set of specific layers for treated group and control group separately. The impact of selection bias in multi-task learning problem can be alleviated via a propensity-dropout regularization scheme~\cite{alaa2017deep}, in which the network is thinned for every training example via a dropout probability that depends on the associated propensity score. The dropout probability is higher for subjects with features that belong in a region of poor overlap in the feature space between treatment and control group.

The Bayesian method also can be extended under multi-task model.  A nonparametric Bayesian method ~\cite{alaa2017bayesian} uses a multi-task Gaussian process with a linear coregionalization kernel as a prior over the vector-valued reproducing kernel Hilbert space. The Bayesian approach allows computing individualized measures of confidence in our estimates via pointwise credible intervals, which are crucial for realizing the full potential of precision medicine. The impact of selection bias is alleviated via a risk-based empirical Bayes method for adapting the multi-task GP prior, which jointly minimizes the empirical error in factual outcomes and the uncertainty in counterfactual outcomes. 

The multi-task model can be extended to multiple treatments even with continuous parameters in each treatment. The dose response network (DRNet) architecture ~\cite{schwab2019learning} with shared base layers, $N_W$ intermediary treatment layers, and $N_W \times E$ heads for the multiple treatment setting with an associated dosage parameter $s$. The shared base layers are trained on all samples, and the treatment layers are only trained on samples from their respective treatment category. Each treatment layer is further subdivided into $E$ head layers. Each head layer is assigned a dosage stratum that subdivides the range of potential dosages $[a_t, b_t]$ into $E$ partitions of equal width $\frac{b-a}{E}$.

\subsection{Meta-Learning Methods}

When designing the heterogeneous treatment effect estimation algorithms, two key factors should be considered: 1) Control the confounders, i.e., eliminate the spurious correlation between the confounder and the outcome; 2) Give an accurate expression of the CATE estimation~\cite{nie2017quasi_rlearner}. The methods mentioned in the previous sections seek to satisfy the two requirements simultaneously, while meta-learning based algorithms separate them into two steps. In general, the meta-learning based algorithms have the following procedures: (1) Estimate the conditional mean outcome $\mathbb{E}[Y|X = x]$, and the prediction model learned in this step is the base learner. (2) Derive the CATE estimator based on the difference of results obtained from step (1). Existing meta-learning methods include T-learner~\cite{kunzel2019metalearners}, S-learner~\cite{kunzel2019metalearners}, X-learner~\cite{kunzel2019metalearners}, U-learner~\cite{nie2017quasi_rlearner} and R-learner~\cite{nie2017quasi_rlearner}, which are introduced in the following. 

In detail, the T-learner~\cite{kunzel2019metalearners} adopts two trees to estimate the conditional treated/control outcomes, which are denoted as $\mu_0(x) = \mathbb{E}[Y(0)|X = x]$ and $\mu_1(x) = \mathbb{E}[Y(1)|X = x]$, respectively. Let $\hat{\mu_0}(x)$ and $\hat{\mu_0}(x)$ denote the trained tree model on the control/treated group. Then the CATE of T-learner estimation is obtained as: $\hat{\tau}_{T}(x) = \hat{\mu_1}(x)-\hat{\mu}_0(x)$. T-learner trains two base models for control and treated groups (the name ``T'' comes from two base model), while S-learner\cite{kunzel2019metalearners} views the treatment assignment as one feature and estimate the combined outcome as: $\mu(x, w) = \mathbb{E}[Y^F|X = x, W = w]$ (The name ``S'' denotes single). $\mu(x, w)$ can be any base model, and we denote the trained model as $\hat{\mu}(x, w)$. The CATE estimator provided by S-learner is then given as: $\hat{\tau}_{S}(x) = \hat{\mu}(x, 1) -\hat{\mu}(x, 0)$.

However, T-learner and S-learner highly rely on the performance of the trained base models. When the number of units in two groups are extremely unbalanced (i.e., the number of one group is much larger than the other), the performance of the base model trained on the small group would be poor. To overcome this problem, X-learner~\cite{kunzel2019metalearners} is proposed, which adopts information from the control group to give a better estimator on the treated group and vice versa. The cross-group information usage is where X-learner comes from, and the X denotes ``cross group''. In detail, X-learner contains three key steps. The first step of X-learner is the same as T-learner, and the trained base learners are denoted as $\hat{\mu}_0(x)$ and $\hat{\mu_1}(x)$. In the second step, X-learner calculates the difference between the observed outcome and the estimated outcome as the imputed treatment effect: In the control group, the difference is the estimated treated outcome subtracts the observed control outcome, denoted as: $\hat{D}_i^C =  \hat{\mu_1}(x) - Y^F$; Similarly, in the treated group, the difference is formulated as: $\hat{D}_i^T = Y^F - \hat{\mu_0}(x)$. After the difference calculation, the dataset is transformed into two groups with imputed treatment effect: control group: $(X_C, \hat{D}^C )$ and treated group: $(X_T, \hat{D}^T )$. On two imputed datasets, the two base learners of treatment effect $\tau_1(x)$( $\tau_0(x)$) are trained with $X_C$($X_T$) as the input and $\hat{D}^C$($\hat{D}^T$) as the output. The last step is to combine the two CATE estimators by weighted average: $\tau_X(x) = g(x)\hat{\tau}_0(x) + (1-g(x))\hat{\tau}_1(x)$, where $g(x)$ is the weighting function ranging from $0$ to $1$. Overall, with the cross information usage and the weighted combination of two CATE base estimator, X-learners can handle the case where the number of units in two groups are unbalanced~\cite{kunzel2019metalearners}.

Different from the regular loss function adopted in X-learner, R-learner, proposed in~\cite{nie2017quasi_rlearner} designs the loss function for CATE estimator based on the Robinson transformation~\cite{robinson1988root_robinson_transformation}. The character ``R'' in R-learner denotes the  Robinson transformation. The Robinson transformation can be derived by rewriting the observed outcome and the conditional outcome: Rewrite the observed outcome as: 
\begin{equation}
 Y_i(W = w_i) = \hat{\mu}_{0}(x_i) + w_i * \tau(x_i) + \epsilon_i(w_i),   
 \label{eqn: robinson_y}
\end{equation}
where $\hat{\mu}_{0}$ is the already-trained control outcome estimator(base learner), $\tau(x_i)$ is the CATE estimator, and $E[\epsilon_i(w_i)|x_i, w_i] = 0$ (under ignorability). The conditional mean outcome can be also rewritten as: 
\begin{equation}
    \hat{m}(x_i) = E[Y|X] = \hat{\mu}_{0}(x_i) + \hat{e}(x_i) * \tau(x_i),
    \label{eqn: robinson_cond_mean}
\end{equation}
where $\hat{e}(x)$ is the already-trained propensity score estimator(base learner). Robinson transformation is obtained by subtracting Eqn.~\eqref{eqn: robinson_y} and Eqn.~\eqref{eqn: robinson_cond_mean}:
\begin{equation}
    Y_i^F-\hat{m}(x_i) = (w_i - \hat{e}(x_i))\tau(x_i) + \epsilon(w_i)
    \label{eqn: robinson_transformation}
\end{equation}
Based on the Robinson transformation, a good CATE estimator should minimize the difference between $Y_i^F-\hat{m}(x_i)$ and $(w_i - \hat{e}(x_i))\tau(x_i)$. Therefore, the objective function of R-learner is as follows:
\begin{equation}
    \tau(\cdot) = \argmin_{\tau}\left\{ \frac{1}{n}\sum_{i = 1}^{n} \left(( Y_i^F - \hat{m}(x_i)) - \left( w_i - \hat{e}(x_i)\right)\tau(x_i) 
    \right)^2 + \Lambda(\tau(\cdot))\right\},
\end{equation}
where $\hat{m}(x_i)$ and $\hat{e}(x_i)$ are pre-trained outcome estimator and propensity score estimator, and $\Lambda(\tau(\cdot))$ is the regularization on $\tau(\cdot)$. 
\section{Methods Relaxing Three Assumptions}
\label{section: relax three assumptions}
In Section~\ref{Section-3}, the causal inference methods based on three assumptions have been introduced in detail, which are the stable unit treatment value assumption (SUTVA), ignorability assumption, and positivity assumption. However, in practice, for some specific applications like social media analysis, which involves dependent network information, special data types (e.g., time series data) or particular conditions (e.g., the existence of unobserved confounders), these three assumptions cannot always hold. In this section, the methods that try to relax certain assumptions will be discussed.

\subsection{Stable unit treatment value assumption (SUTVA) assumption}

Stable Unit Treatment Value Assumption (SUTVA) states that the potential outcomes for any unit do not vary with the treatment assigned to other units, and, for each unit, there are no different forms or versions of each treatment level, which lead to different potential outcomes. This assumption mainly focuses on two aspects: (1) Units are independent and identically distributed (i.i.d.); (2) there only exists a single level for each treatment.  An extensive literature exists on making causal inferences under SUTVA, but when considering many real-world situations, it may not always be the case. In the following, SUTVA will be discussed from these two aspects.
 
The assumption of independent and identically distributed samples is ubiquitous in most causal inference methods, but this assumption cannot hold in many research areas, such as social media analytics ~\cite{guo2019learning}~\cite{shalizi2011homophily}, herd immunity, and signal processing ~\cite{volodymyr2015human} ~\cite{sutskever2014sequence}. Causal inference in non-i.i.d. contexts is challenging due to the presence of both unobserved confounding and data dependence. For example, in social networks, subjects are connected and influenced by each other. 

For such network data, SUTVA cannot hold anymore. Under this situation, instances are inherently interconnected with each other through the network structure and hence their features are not independent identically distributed samples drawn from a certain distribution. Applying Graph Convolutional Networks into causal inference model is an approach to handle the network data~\cite{guo2019learning}. In particular, the original features of subjects and the network structure are mapped to a representation space, in order to get the representation of confounders. Furthermore, the potential outcomes could be inferred using treatment assignments and confounder representations.

The dependence in data often leads to interference because some subjects' treatments can affect others' outcomes~\cite{hudgens2008toward, ogburn2014causal}. This difficulty can impede the identification of causal parameters of interest. Extensive work has been developed on identification and estimation of causal parameters under interference~\cite{pena2018reasoning, hudgens2008toward, ogburn2014causal,tchetgen2012causal}. For this problem, a strategy proposed by ~\cite{sherman2018identification_dependent_data} is to use segregated graphs ~\cite{shpitser2015segregated}, a generalization of latent projection mixed graphs ~\cite{verma1991equivalence}, to represent causal models.
 
Modeling time series data is another important problem in causal inference, which does not satisfy the independent and identically distributed assumption. Most of the existing methods use regression models for this problem but the accuracy of inference depends greatly on whether the model fits the data. Therefore, selecting a right and appropriate regression model is of crucial importance, but in practice, it is not easy to find the perfect one. \cite{chikahara2018causal} proposes a supervised learning framework that uses a classifier to replace regression models. It presents a feature representation that employs the distance between the conditional distributions given past variable values and shows experimentally that the feature representation provides sufficiently different feature vectors for time series with different causal relationships. For the time series data, another issue that needs to be considered is hidden confounders. A time series deconfounder ~\cite{bica2019time} was developed, which leverages the assignment of multiple treatments over time to enable the estimation of treatment effects even in the presence of hidden confounders. This time series deconfounder uses a recurrent neural network architecture with multitask output to build a factor model over time and infer substitute confounders, which render the assigned treatments conditionally independent. Then it performs causal inference using the substitute confounders. 

For the second direction in SUTVA assumption, it assumes that there only exists one version for each treatment. However, if adding one continuous parameter into the treatment, this assumption cannot hold anymore. For example, estimating individual dose response curves for a couple of treatments requires adding an associated dosage parameter (categorical or continuous) for each treatment. Under this situation, for each treatment, it will have multiple versions for categorical dosage parameters or infinite versions for continuous dosage parameters. One way to solve this problem is to convert the continuous dosage into a categorical variable and then treat every medication with specific dosage as one new treatment, so that it will satisfy the SUTVA assumption again~\cite{schwab2019learning}.

Another example that breaks the SUTVA is the dynamic treatment regime, which consists of a sequence of decision rules, one per stage of intervention ~\cite{chakraborty2013statistical}. One useful application of dynamic treatment is precision medication. It includes more individualization to adjust which type of treatment should be used, or how many the dosage is best in response to the patient's background characteristics, the illness severity and other heterogeneity, aiming to get the optimal treatment strategy. These heterogeneities are called tailoring variables. To get a useful dynamic treatment regime, \cite{lavori2000design} introduces one 'biased coin adaptive within-subject' (BCAWS) design. Then, \cite{murphy2005experimental} presents one general framework of this type of design, which uses sequential multiple assignment randomized trials (SMART) for developing decision rules in that each individual may be randomized multiple times and the multiple randomizations occur sequentially over time. 

For estimating optimal dynamic decision rules from observational data, Q~\cite{watkins1989learning,watkins1992q} and A  ~\cite{robins2004optimal, murphy2003optimal} learning are two main approaches for estimating the optimal dynamic treatment regime. Q in Q-learning denotes ``quality''. Q-learning is a model-free reinforcement learning algorithm, which employs posited regression models for estimating outcome at each decision point given units' information. In advantage learning (A-learning), models are posited only for the part of the regression including contrasts among treatments and for the probability of observed treatment assignment at each decision point, given units' information. Both methods are implemented through a backward recursive fitting procedure that is related to dynamic programming ~\cite{bather2000decision}.

\subsection{Unconfoundedness assumption}
\label{subsection: Unconfoundedness assumption}

The ignorability assumption is also named as unconfoundedness assumption. Given the background variable, $X$, the treatment assignment $W$ is independent to the potential outcomes, i.e., $W \independent {Y(W = 0), Y(W = 1)} | X$. With this unconfoundedness assumption, for the units with the same background variable $X$, their treatment assignment can be viewed as random. Obviously, identifying and collecting all of background variables is impossible, and this assumption is very difficult to satisfy. For example, in an observational study that tries to estimate the individual treatment effect of a medicine, instead of randomized experiments, the medicine is assigned to individuals based on a series of factors. Some factors (e.g., socioeconomic status) are challenging to measure and therefore become hidden confounders. Existing work overwhelmingly relies on the unconfoundedness assumption that all confounders can be measured. However, this assumption might be untenable in practice. In the above example, units' demographic attributes, such as their home address, consumption ability or employment status, may be the proxies for socioeconomic status. Leveraging big data, it is possible to find a proxy for the latent and unobserved confounders. 

Variational autoencoder has been used to infer the complex non-linear relationships between the observed confounders and joint distribution of the latent confounders, treatment assignment and outcomes~\cite{louizos2017causal}. The joint distribution of the latent confounders and the observed confounders can be approximately recovered from the observations. An alternative way is to capture their patterns and control their influence by incorporating the underlying network information. Network information is also a reasonable proxy for the unobserved confounding. \cite{guo2019learning} applies GCN on network information to get the representation of hidden confounders. Moreover, in ~\cite{guo2019counterfactual}, graph attention layers are used to map the observed features in networked observational data to the D-dimensional space of partial latent confounders, by capturing the unknown edge weights in the real-world networked observational data.

An interesting insight mentioned in \cite{veitch2019using} is that, even if the confounders are observed, it doesn't mean all the information they contain is useful to infer the causal effect. Instead, requiring the part of confounders actually used by the estimator is sufficient. Therefore, if a good predictive model for the treatment can be built, one may only need to plug the outputs into a causal effect estimate directly, without any need to learn the whole true confounders. In \cite{veitch2019using}, the main idea is to reduce the causal estimation problem to a semi-supervised prediction of both the treatments and outcomes. Networks admit high-quality embedding models that can be used for this semi-supervised prediction. In addition, embedding methods can also offer an alternative to fully specified generative models. 

Only using observational data to solve the confoundings problem is always difficult. Another way is to combine the experimental data and observational data together. In ~\cite{kallus2018removing}, limited experimental data is used to correct the hidden confounding in causal effect models trained on larger observational data, even if the observational data do not fully overlap with the experimental data. This method makes strictly weaker assumptions than existing approaches.

For estimating treatment effects from longitudinal observational data, existing methods usually assume that there are no hidden confounders. This assumption is not testable in practice and, if it does not hold, leads to biased estimates. \cite{bica2019time} infers substitute confounders that render the assigned treatments conditionally independent. Then it performs causal inference using the substitute confounders. This method can help estimate treatment effects for time series data in the presence of hidden confounders.

Above methods all aim to solve the problems about the observed and unobserved confounders. Are there any other ways to get around the unconfoundedness assumption and conduct causal inference? One way is to use instrumental variables that only affect treatment assignment but not the outcome variable. Changes in the instrumental variables would lead to the different assignment of treatment, which is independent of the latent variables, and this assignment is as good as randomization for the purposes of causal inference. \cite{hartford2017deep} broke instrumental variables analysis into two supervised stages that can each be targeted with deep networks. It models the conditional distribution of the treatment variable given the instruments and covariates, and then employs a loss function involving integration over the conditional treatment distribution. The deep instrumental variables framework also takes advantage of existing supervised learning techniques to estimate causal effects.

\subsection{Positivity assumption}
The positivity assumption, also known as covariate overlap or common support, is a necessary assumption for the identification of treatment effect in the observational study. However, little literature discusses the satisfaction of this assumption in the high dimensional datasets. \cite{d2017overlap} argues that the positivity assumption is a strong assumption and is more difficult to be satisfied in the high-dimensional datasets. To support the claim, the implication of the strict overlap assumption is explored, and it shows that strict overlap restricts the general discrepancies between the control and treated covariates. Therefore, the positivity assumption is stronger than the investigator expected. Based on the above implication, methods that eliminate the information about the treatment assignment while still hold the unconfounderness assumption are recommended, such as trimming~\cite{rosenbaum1983central,petersen2012diagnosing_trimming,crump2009limited_overlap} that drops the records in the region without overlap, and instrumental variable adjustment methods~\cite{myers2011effects_instrumental,pearl2012bias-amplifying,ding2017instrumental} that eliminate the instrumental variables from covariates.

\section{Guideline about Experiment}
\label{sec:experiment}
In this section, we provide the related experimental information, including the available datasets that are commonly adopted in the experiments, and the open-source codes of the methods mentioned in the previous two sections. 

\subsection{Available Datasets}
\subsubsection{Datasets for Section~\ref{Section-3}} 
Because the counterfactual outcome can never be observed, it's hard to find the dataset that perfectly satisfies the requirements of the experiment that it is an observational dataset with the ground truth ATE (or ITE) available. The datasets used in the literature are often semi-synthetic datasets. Some datasets, such as IHDP dataset, are obtained from the randomized dataset by generating their observed outcome according to a certain generation process and removing a biased subset to mimic the selection bias in the observational dataset. Some datasets, such as Jobs dataset, combine the randomized dataset and the observational control dataset together to create the selection bias. The details of the available benchmark datasets are in the following. 

\textbf{IHDP}. This dataset is a commonly adopted benchmark dataset. This dataset is generated based on the randomized controlled experiment conducted by Infant Health and Development Program~\cite{brooks1992ihdp}, whose targets are low-birthweight, premature infants.  The pre-treatment covariates are $25$ variables measuring the aspects about the children and their mothers, such as birth weight, head circumference, neonatal health index, prenatal care, mother's age, education, drugs, alcohol, etc. In the treated group, the infants are provided with both intensive high-quality childcare and specialist home visits~\cite{hill2011bayesian}. The outcome is the infants' cognitive test score and can be simulated through the NPCI package\footnote{\url{https://github.com/vdorie/npci}}. Besides, a biased subset of the treated group is required to be removed to simulate the selection bias. An example of IHDP dataset whose outcome is simulated by the setting ``A'' of NPCI package can be downloaded from \url{http://www.mit.edu/~fredrikj/files/ihdp_100.tar.gz}.

\textbf{Jobs}. The jobs datasets used in the observational study~\cite{dehejia1999causal_jobs,dehejia2002propensity_jobs,shalit2017estimating} is the combination of Lalonde experiment data and the PSID comparison group. Both Lalonde and PSID datasets can be downloaded from NBER website\footnote{\url{http://users.nber.org/~rdehejia/data/nswdata2.html}}. The pre-treatment covariates are $8$ variables such as age, education, ethnicity, as well as earnings in 1974 and 1975. The people in the treated group take part in the job training while in the control group are not. The outcome is employment status. 

\textbf{Twins}. Twins dataset is first introduced in~\cite{louizos2017causal} and is adopt by various observational studies~\cite{louizos2017causal,yoon2018ganite,yao2018representation}. Twins dataset is constructed on the data of twins birth in the USA between 1989-1991~\cite{almond2005costs}\footnote{\url{www.nber.org/data/linked-birth-infant-death-data-vital-statistics-data.html}}. In~\cite{louizos2017causal}, the twins whose gender is the same and weight is less than $2000g$ are selected into records. For each twin pair, there are in total $40$ pre-treatment covariates measuring the pregnancy, twins birth and the parents, such as the number of gestation weeks before birth, the quality of care during pregnancy, pregnancy risk factors (Anemia, alcohol use, tobacco use, etc.), adequacy of care, residence and so on. The outcome is the one-year mortality. The treatment is being the heavier one in the twins, and the outcome is the one-year mortality. In twins dataset, both treated (the heavier one in the twin) and control (the lighter one in the twin) outcomes are observed. The treatment assignment usually defined by the users to simulate the selection bias. For example, in~\cite{louizos2017causal,yoon2018ganite}, the selection bias is created by the following procedure: $W_i|\mathbf{X}_i \sim Bern(Sigmoid(\mathbf{w}^{'}\mathbf{X}_i)+n)$, where $\mathbf{w} \sim U(-0.1,0.1)^{40 \times 1}$ and $n\sim \mathbf{N}(0,0.1)$.

\textbf{ACIC datasets}. Starting from 2016, every year, the Atlantic Causal Inference Conference holds the causal inference data analysis challenge, which provides some datasets targeting different causal inference problems.

2016 Challenge: The goal of ACIC 2016 Challenge is to better understand which approaches to causal inference perform well in particular observational study settings\footnote{\url{https://jenniferhill7.wixsite.com/acic-2016/competition}}.
The datasets contain 77 datasets with varying degrees of non-linearity, sparsity, correlation between treatment assignment and outcome, non-linearity of treatment effect, overlapping. The covariates are real-world data from the Infant Health and Development Program dataset~\cite{brooks1992ihdp}, which consists of 58 variables. The treatment, factual outcome, and counterfactual outcome are all generated by simulation, and the selection bias is created by removing treated children with non-white moms. The whole datasets can be downloaded from \url{https://drive.google.com/file/d/0B7pG5PPgj6A3N09ibmFwNWE1djA/view}, and the summarization of this year's challenge is in~\cite{dorie2019automated}.

2017 Challenge: ACIC 2017 challenge focused on the estimation and inference for conditional average treatment effects (CATEs) in the presence of targeted selection. Targeted selection means the likelihood that an individual receives treatment is a function of the expected response of that individual if left untreated, which leads to strong confounding~\cite{hahn2019atlantic17}. The same as the previous year's challenge, the covariates are from Infant Health and Development Program dataset~\cite{brooks1992ihdp}, but only $8$ variables are used. The outcomes and the treatment assignments are generated according to $32$ distinct, fixed, data generating processes representing four different types of errors. For every data generating process, 250 independent replicate data sets were produced, and overall, there are a total of $8,000$ data sets.

2018 Challenge: The ACIC 2018 challenge has two different tasks focusing on two sub-challenges: censoring and scaling. Censoring means some of the samples may not have observed outcomes. Therefore, the dataset used by censoring challenges contains missing outcome values for some of the samples. The dataset for scaling challenge contains $48$ datasets whose data sizes, and they are not censored. The details of the above datasets are available at~\url{https://www.synapse.org/#!Synapse:syn11294478/wiki/486304}

2019 Challenge: This challenge focuses on estimating the ATE on the quasi real-world dataset with low dimensional data and high dimensional data\footnote{\url{https://sites.google.com/view/acic2019datachallenge/data-challenge?authuser=0}}. The datasets for this challenge contain several datasets with different variables size and record size, and the R code for data generation is available at \url{https://drive.google.com/file/d/1Qqgmb3R9Vt9KTx6t8i_5IbFenylsPfrK/view}.

\textbf{IBM causal inference benchmark}.
This dataset is created in~\cite{2018_CausalBenchmark} and is available at \url{https://github.com/IBM-HRL-MLHLS/IBM-Causal-Inference-Benchmarking-Framework}.
This dataset uses the cohort of 100K samples in Linked Births and Infant Deaths Database (LBIDD){\footnote{\url{https://www.cdc.gov/nchs/nvss/linked-birth.htm}}} as the fundamental set of covariates. The treatment, factual outcome, and the counterfactual outcome are generated by simulation. 

\textbf{BlogCatalog}. This dataset is used for causal inference with networked observational data~\cite{guo2019learning}. It is a social blogger network. A blogger is one observation. The bloggers are connected by some social relationships in this dataset. The features are bag-of-words representations of keywords in bloggers' descriptions. The outcomes are the opinions of readers on each blogger. A blogger belongs to the treated group (control group) if her blogs are read more on mobile devices (desktops). 

\textbf{Flickr}. This dataset includes networked observational data in~\cite{guo2019learning}. 
Flickr is a photo-sharing platform and social network where users upload photos for others to see. In this dataset, the users with Flickr account are observations, and the users are connected by some social relationships. The features of each user are the tags of interest. The outcomes and treatment assignment are the same as BlogCatalog.

\textbf{News}.  The News benchmark includes 5000 randomly sampled news articles from the NY Times corpus. It contains the data on the opinion of media consumers on news items and was originally introduced as a benchmark for counterfactual inference in the setting with two treatment options ~\cite{johansson2016learning}. It can be extended to multiple treatments with associated dosage parameters ~\cite{schwab2019learning}. The details can be found in \url{https://archive.ics.uci.edu/ml/datasets/bag+of+words}.

\textbf{MVICU}. The Mechanical Ventilation in the Intensive Care Unit (MVICU) benchmark is used to estimate individual dose-response curves for a couple of treatments with an associated dosage parameter ~\cite{schwab2019learning}. This dataset includes patients' responses to different configurations of mechanical ventilation in the intensive care unit. The data was sourced from the publicly available MIMIC III database which documents a diverse and very large population of ICU patient stays and contains comprehensive and detailed clinical data. ~\cite{saeed2011multiparameter}.

\textbf{TCGA}. The Cancer Genome Atlas (TCGA) is the world's largest and richest collection of genomic data. This dataset is used to estimate individual dose-response curves for a couple of treatments with an associated dosage parameter ~\cite{schwab2019learning}. The TCGA project collected gene expression data from various types of cancers in 9659 individuals ~\cite{weinstein2013cancer}. The treatment options are medication, chemotherapy and surgery. The outcome is the risk of cancer recurrence after receiving the treatment. TCGA data (controlled access and open access data) are available via the Genomic Data Commons (GDC) \url{https://gdc.cancer.gov/}.

\textbf{Saccharomyces cerevisiae (yeast) cell cycle gene expression dataset}. This is one time series dataset. A time series with the length T = 57 was created by combining four short time series that were measured in different microarray experiments~\cite{chikahara2018causal}. 

\textbf{THE}. The Tennessee Student/Teacher Achievement Ratio (STAR) experiment is a four-year longitudinal class-size study funded by the Tennessee General Assembly and conducted by the State Department of Education to measure the influence of class size  (small class, regular class and regular-with-aide class) on student achievement tests and non-achievement measures ~\cite{achilles2008tennessee}. Because this is one randomized controlled experiment, CATE estimates are unbiased due to unconfoundedness. Confounders are artificially introduced by selectively removing a biased subset of samples~\cite{kallus2018removing}.

\textbf{FERTIL2}. This dataset aims to study the impact of more than or exactly 7 years of education for a woman on the number of children in the family~\cite{wooldridge2010econometric}. Several observed confounders are included in the dataset, such as age, whether the family has a TV, whether the woman lives in the city. The instrumental variable is a binary indicator of whether the woman was born in the first half of the year. This dataset is used for research about instrumental variables~\cite{ding2017instrumental}.

\subsection{Codes/Packages}
In this part, we summarize the available codes or tool-boxes for causal inference. The codes for methods that mentioned in Section~\ref{Section-3} are provided in Table~\ref{tab:sec3 tool-box} and Table~\ref{tab: sec3 codes}, where Table~\ref{tab:sec3 tool-box} lists the tool-boxes with their supported methods and languages, and Table~\ref{tab: sec3 codes} lists the open-source code of one specific method.

\begin{table}[ht]
    \centering
        \caption{Available Tool-boxes for Causal Inference }
    \begin{tabular}{|c|c|c|l|}
    \toprule
     \textbf{Tool-box}    & \textbf{Supporting methods} & \textbf{Language} & \textbf{Link}\\
     \midrule
      \multirow{ 2}{*}{Dowhy~\cite{dowhy}}   & Propensity-based Stratification,
 & \multirow{ 2}{*}{Python} &  \multirow{ 2}{*}{\url{https://github.com/microsoft/dowhy}}\\
&PSM,
IPW, Regression&&\\
\midrule
\multirow{ 2}{*}{Causal ML} &Tree-based algorithms,  & \multirow{ 2}{*}{Python}& \multirow{ 2}{*}{\url{https://github.com/uber/causalml}}\\
&X/T/X/R-learner&&\\
    \midrule
    \multirow{ 4}{*}{EconML~\cite{econml}} &Doubly Robust Learner,  & \multirow{ 4}{*}{Python} & \\
    &  Orthogonal Random Forests, & &\href{https://github.com/microsoft/EconML#blogs-and-publications}{https://github.com/microsoft/} \\
    &Meta-Learners, &&\href{https://github.com/microsoft/EconML#blogs-and-publications}{EconML\#blogs-and-publications}\\
    &Deep Instrumental Variables&&\\
    \midrule
    \multirow{3}{*}{causalToolbox} & BART, Causal Forest, &  \multirow{3}{*}{R} &\href{https://github.com/soerenkuenzel/causalToolbox}{https://github.com/soerenkuenzel} \\
    & T/X/S-learner with  & &\href{https://github.com/soerenkuenzel/causalToolbox}{/causalToolbox} \\
    &BART/RF as base learner&&\\
    \bottomrule
    \end{tabular}
    \label{tab:sec3 tool-box}
\end{table}

\begin{longtable}{|c|c|l|}
\caption{Available Codes of Methods in Section~\ref{Section-3}} \label{tab: sec3 codes} \\
\hline \multicolumn{1}{|c|}{\textbf{Method}} & \multicolumn{1}{c|}{\textbf{Language}} & \multicolumn{1}{c|}{\textbf{Link}} \\ \hline 
\endfirsthead
\multicolumn{3}{c}%
{{\bfseries \tablename\ \thetable{} -- continued from previous page}} \\
\hline \multicolumn{1}{|c|}{\textbf{Method}} & \multicolumn{1}{c|}{\textbf{Language}} & \multicolumn{1}{c|}{\textbf{Link}} \\ \hline 
\endhead
\hline 
\multicolumn{3}{|r|}{{Continued on next page}} \\ \hline
\endfoot
\hline \hline
\endlastfoot
IPW & R & \href{https://cran.r-project.org/web/packages/ipw/index.html}{https://cran.r-project.org/web/packages/ipw/index.html}\\
     \midrule
     \multirow{2}{*}{DR} & \multirow{2}{*}{R} & fastDR: \url{https://github.com/gregridgeway/fastDR}
     \\
     && DR for High dimension: \url{https://github.com/gregridgeway/fastDR}\\
     \midrule
     Principal Stratification & R &
     \url{https://cran.r-project.org/web/packages/sensitivityPStrat/index.html}\\
     \midrule
     Stratification &R & \url{https://cran.r-project.org/web/packages/stratification/} \\
     \midrule
     PSM & \multirow{3}{*}{Python}&\multirow{3}{*}{\url{https://cran.r-project.org/web/packages/PSW/}}\\
     overlap weight&&\\
     trapezoidal weight&&\\
     \midrule
     Matching based Alg.: & \multirow{6}{*}{R} & \multirow{6}{*}{\url{https://cran.r-project.org/web/packages/MatchIt/}} \\
      exact matching, &&\\full matching, &&\\  genetic matching, &&\\ nearest neighbor matching,&&\\ optimal matching, &&\\subclassification &&\\\midrule
     PSM & Python& \url{https://github.com/akelleh/causality}\\
     \midrule
     Perfect Match & Python&\url{https://github.com/d909b/perfect_match} \\
     \midrule
     optimal matching & R & \url{https://cran.r-project.org/web/packages/Matching/}\\\midrule
      CEM & R & \url{https://cran.r-project.org/web/packages/cem/}\\\midrule
      TMLE &  R & \href{https://cran.r-project.org/web/packages/tmle/index.html}{https://cran.r-project.org/web/packages/tmle/index.html}\\\midrule
      \multirow{3}{*}{CMGP~\cite{alaa2017bayesian}}& \multirow{3}{*}{Python} &  \href{https://bitbucket.org/mvdschaar/mlforhealthlabpub/src/baa0aa33a6af3fe490484c9e11e3a158968ae56a/alg/causal_multitask_gaussian_processes_ite/}{https://bitbucket.org/mvdschaar/mlforhealthlabpub/src/} \\
      &&\href{https://bitbucket.org/mvdschaar/mlforhealthlabpub/src/baa0aa33a6af3fe490484c9e11e3a158968ae56a/alg/causal_multitask_gaussian_processes_ite/}{baa0aa33a6af3fe490484c9e11e3a158968ae56a/}\\
      &&\href{https://bitbucket.org/mvdschaar/mlforhealthlabpub/src/baa0aa33a6af3fe490484c9e11e3a158968ae56a/alg/causal_multitask_gaussian_processes_ite/}{alg/causal\_multitask\_gaussian\_processes\_ite/} \\\midrule
      \multirow{2}{*}{BART} & R  &\url{https://cran.r-project.org/web/packages/BayesTree/index.html}\\ & Python & \url{https://github.com/JakeColtman/bartpy}\\ 
     \midrule
     \multirow{2}{*}{GANITE~\cite{yoon2018ganite}} & Python &\href{https://bitbucket.org/mvdschaar/mlforhealthlabpub/src/baa0aa33a6af3fe490484c9e11e3a158968ae56a/alg/ganite/}{https://bitbucket.org/mvdschaar/mlforhealthlabpub/src/} \\
     & &\href{https://bitbucket.org/mvdschaar/mlforhealthlabpub/src/baa0aa33a6af3fe490484c9e11e3a158968ae56a/alg/ganite/}{baa0aa33a6af3fe490484c9e11e3a158968ae56a/alg/ganite/}\\
     \midrule
      BNN~\cite{johansson2016learning}, & \multirow{3}{*}{Python} & \multirow{3}{*}{\url{https://github.com/clinicalml/cfrnet}}\\ CFR-MMD~\cite{shalit2017estimating},&&\\ CFR-WASS~\cite{shalit2017estimating}&&\\
      \midrule
      CEVAE &Python &\url{https://github.com/AMLab-Amsterdam/CEVAE}\\
      \midrule
      SITE~\cite{yao2018representation} & Python &  \url{https://github.com/Osier-Yi/SITE}\\
      \midrule
       grf & R &\url{https://cran.r-project.org/web/packages/grf/index.html}\\
       \midrule
       R-learner & R &  \url{https://github.com/xnie/rlearner/blob/master/R/xlearner.R}\\\midrule
       Residual Balancing & R & \url{https://github.com/swager/balanceHD}\\ 
        \midrule
        CBPS & R &  \url{https://github.com/kosukeimai/CBPS}\\
        \midrule
        dragonnet & Python &  \url{github.com/claudiashi57/dragonnet}\\
        \midrule
         Entropy Balancing & R &  \url{https://cran.r-project.org/web/packages/ebal/}\\
    
        \midrule
        DRNets~\cite{schwab2019learning} & Python &  \url{https://github.com/d909b/drnet}\\
        \midrule
        Network Deconfounder~\cite{guo2019learning} & Python &  \url{https://github.com/rguo12/network-deconfounder-wsdm20}\\
        \midrule
        Network Embeddings ~\cite{veitch2019using} & Python &  \url{https://github.com/vveitch/causal-network-embeddings}\\
        \midrule
        RMSN~\cite{lim2018forecasting} & Python &  \url{https://github.com/vveitch/causal-network-embeddings}\\
        \midrule
        TMLE~\cite{petersen2014targeted} & R &  \url{https://github.com/joshuaschwab/ltmle}\\
        \midrule
        LCVA~\cite{rakesh2018linked} & Python &  \url{https://github.com/rguo12/CIKM18-LCVA}\\
\end{longtable}

\section{Applications}
\label{sec: application}

Causal inference has a variety of applications in real-world scenarios. In general, the applications of causal inference can be categorized into three directions: 
\begin{itemize}
    \item [(1)] Decision evaluation. This is a natural application of treatment effect estimation as it is consistent with the objective of treatment effect estimation.
    \item[(2)] Counterfactual estimation. Counterfactual learning greatly helps the areas related to decision making, as it can provide the potential outcomes of different decision choices (or policies).
    \item[(3)] Dealing with selection bias. In many real-world applications, the records appear in the collected dataset are not representative of the whole population that is interested. Without appropriately handling the selection bias, the generalization of the trained model would be hurt. 
\end{itemize}
In this section, we will discuss how causal inference benefits various real-world applications in detail.

\subsection{Advertising}
Properly measuring the effect of an advertising campaign can answer critical marketing questions such as whether a new advertisement increases the clicks, or whether a new campaign increases sales, etc. Since conducting randomized experiments is expensive and time-consuming, estimating the advertisement effect from the observational data is attracting increasing attention in both industry and research communities ~\cite{sun2015causal_online_advertising,wang2015robust_complex_ad}. In~\cite{li2016matching_digital_ijcai16}, the randomized nearest neighbor matching method is proposed to estimate the treatment effect of digital marketing campaigns. In~\cite{fong2018covariate_CBGPS}, the covariate balancing generalized propensity score (CBGPS), discussed in Section~\ref{subsection: sample re-weighting}, is applied to analyze the efficacy of political advertisements. 

However, in the online advertisement area, it is often required to deal with complex advertisement treatments, which could be a discrete or continuous, uni- or multi-dimensional treatment~\cite{sun2015causal_online_advertising}. One advertisement is set as the baseline treatment, and the treatment effect is obtained by comparing the potential outcome of the treatment with different values and the baseline treatment. To estimate the potential outcome of treatment with multi-dimensional values, tree-based method~\cite{wang2015robust_complex_ad} and sparse additive model based method~\cite{sun2015causal_online_advertising} are proposed to enable the comparison between potential treatments and the baseline treatment. 

In addition to purely observational data, in the real-world scenarios, it is often the case that dataset is comprised of large samples from control condition(i.e., the old treatment) and small samples (possibility unrepresentative) from a randomized trial which contains both the control condition and the new treatment. In~\cite{rosenfeld2017large_samll}, the small randomized trial dataset is connected with the large control dataset using the minimal set of modeling assumptions, which implies the models to predict the control and treated outcome to be similar. Under this assumption, the proposed method jointly learns the control and treated outcome predictor and regularizes the difference between the parameters of two predictors. 

The above discussions show the potential applications of the treatment effect estimation in decision evaluation: measuring the effect of the advertisement campaign. Another important application is to handle the selection bias. Due to the existing selection mechanism in the advertising systems, there is a distribution discrepancy between the displayed and non-displayed events~\cite{ad_click}. Ignoring such bias would make the advertisement click prediction inaccurate, which would cause a loss of revenue. To handle the selection bias, similar to the doubly robust estimation mentioned in Section~\ref{subsection: sample re-weighting}, doubly robust policy learning is proposed in~\cite{dudik2011doubly_policy}. It contains two sub-estimators: direct method estimator obtained from the observed samples, and IPS estimator with the propensity score as the sample weight.

Furthermore, some works notice the difficulty of propensity score estimation due to the deterministic advertisement display policy in commercial advertisement systems. If the display policy is stochastic, the advertisements with low propensity scores still have a chance to appear in the observational dataset so that IPS can correct the selection bias. However, when the display policy is deterministic, the advertisements with low propensity scores are always absent in the observation, which makes propensity score estimation failed. This challenges motives the work of propensity-free doubly robust method proposed in~\cite{ad_click} which improves the original doubly robust method in two folds: 
(1) Train the direct method on a small but unbiased data obtained under the uniform policy, which, to a certain degree, prevents the selection bias propagating to the non-displayed advertisements.
(2) Avoid the propensity score estimation by setting the propensity score of the observed items as $1$ and combines IPS with the direct method. In a nutshell, this propensity score free method relies on the direct method trained on a small unbiased dataset to give an unbiased prediction of the advertisement click. 

Apart from the applications discussed above, another important application is the advertisement recommendation, which is merged into the next subsection.

\subsection{Recommendation}
The recommendation is highly correlated with the treatment effect estimation, as exposing the user to an item in the recommendation system can be viewed as applying one specific treatment to a unit~\cite{schnabel2016recommendations, lada2018observational_rec}. Similar to the dataset used in the treatment effect estimation, the dataset used in the recommendation are usually biased due to the self-selection of the users. For example, in the movie rate dataset, users tend to rate the movies that they like: the horrible movie ratings are mostly made by horror movie fans and less by romantics movie fans. Another example is the advertisement recommendation datasets. The recommendation system would only recommend the advertisements to the users whom the system believes are interested in those advertisements. In the above examples, the records in the datasets are not representative of the whole population, which is the selection bias. The selection bias brings challenges to both recommendation model training and evaluation. Re-weighting samples based on the propensity score is a powerful method to solve the problems that stem from selection bias. The improved performance estimation after propensity score weighting can be calculated as follows: 
\begin{equation}
    \hat{R}_{\text{IPS}}(\hat{Y}|P) = \frac{1}{U \cdot I} \sum_{(u,i): O_{u,i} = 1}\frac{\delta_{u,i}(Y, \hat{Y})}{P_{u,i}},
\label{eqn: IPS_recommendation}
\end{equation}
where $\hat{Y}$ is the value upon which to measure the quality of a recommendation system, $U$ is the number of users, and $I$ is the number of items. $O_{u,i}$ is a binary variable to indicates the interaction of the $u$-th user with the $i$-th item in the observational data. $\delta_{u,i}(\cdot, \cdot)$ can be any classical quality measure of a recommendation, such as cumulative gain(CG), discounted cumulative gain (DCG), and precision at $k$.
$P$ is the marginal probability matrix, whose entry is defined as $P_{u,i} = P(O_{u,i} = 1)$. The improved quality measure is an unbiased estimation to the real measurement $R(\hat{Y})$ over the whole population, which is defined as $R(\hat{Y}) = \frac{1}{U \cdot I} \sum_{u = 1}^{U}\sum_{i = 1}^{I}\delta_{u,i}(Y, \hat{Y})$. Based on the unbiased quality measurement, in~\cite{schnabel2016recommendations}, the propensity-score empirical risk minimization (ERM) for recommendation is proposed: $\hat{Y} \in \mathcal{H}$ is selected to optimize the following problem: $\hat{Y}^{ERM} = \text{argmin}_{\hat{Y} \in \mathcal{H}} \{ \hat{R}_{\text{IPS}}(\hat{Y}|P)\}$, where $\hat{R}_{\text{IPS}}(\hat{Y}|P)$ is defined in Eqn.~\eqref{eqn: IPS_recommendation}. Later, various works are developed to drawbacks of propensity score weighting, including estimation variance reduction~\cite{swaminathan2017off_IPS,saito2019eliminating_IPS}, handle data sparsity~\cite{swaminathan2017off_IPS,saito2019eliminating_IPS}, doubly robust estimation~\cite{wang2019doubly_recommendation}.

In addition to using IPS or doubly robust estimation based method to overcome the selection bias, similar to the advertisement domain, some works also adopt a small unbiased dataset to correct selection bias. In this case, the dataset contains a large set of logged feedback records under control policy and a small set of records under the randomized recommendation. CausalEmbed(CausE)~\cite{bonner2018causal_emb} is a representative method in this direction, which proposed a new matrix factorization algorithm. In detail, CausalEmbed jointly factorizes the matrix of those two datasets and links these two models by regularizing the difference between treated and control representation.

\subsection{Medicine}
Learning the optimal per-patient treatment rules is one of the promising goals of applying treatment effect estimation methods in the medical domain. When the effect of different available medicines can be estimated, doctors can give a better prescription accordingly. In~\cite{shalit2019can-individual}, two challenges are mentioned to fulfill such goal: the existence of confounders and the existence of unobserved confounders. Although analyzing from the randomized experimental dataset is the golden solution, it has the following limitations: (1) The goal of randomized experimental data is to analyze the ATE instead of ITE, so the data size is often small, which limits the capacity to derive personalized treatment rules. (2) As mentioned in section~\ref{sec2}, conducting randomized trials is often expensive, time-consuming, and sometimes maybe unethical. Therefore, deriving personalized treatment rules from the observational dataset or the combination of experimental and observational data are two fruitful directions~\cite{shalit2019can-individual}. 

For the direction of utilizing observational dataset, various methods derive the personalized treatment rules guided by the estimated ITE under the unconfoundedness assumption, such as deep-treat~\cite{atan2018deep_treat}, tiered case-cohort design based method~\cite{kessler2019machine_precision}. However, in this area, there are limited works to handle the unobserved confounders, and methods discussed in Section~\ref{subsection: Unconfoundedness assumption} have great potentials to explore.

\subsection{Reinforcement Learning}
From the perspective of reinforcement learning, ITE estimation can be viewed as a contextual multi-armed bandit problem with the treatment as the action, the outcome as the reward, and the background variables as the contextual information. Arm exploration and exploitation is similar to randomized trials and observational data. Therefore, these two areas share some similar critical challenges: (1) How to get an unbiased outcome/reward estimation? (2) How to handle either the observed or unobserved confounders that affect both the treatment assignment/action choice and the outcome/reward? 

To obtain an unbiased reward estimation, importance sampling weighting~\cite{precup2000eligibility} is the common method adopted in the offline policy evaluation. The weight is set as the probability between the target policy and the logged (observed) policy, which is analogous to IPW mentioned in Section~\ref{subsection: sample re-weighting}. However, importance sampling proposed in~\cite{precup2000eligibility} suffers from high variance and highly relies on the assigned weights. To improve this, similar to the doubly robust method in ATE estimation, doubly robust policy evaluation is proposed in ~\cite{dudik2011doubly_policy}. Later, various methods~\cite{swaminathan2017off_IPS,zou2019context_balancing_offline_PE,atan2018counterfactual_policy,bietti2018practical,tennenholtz2019off,li2012unbiased,kallus2019intrinsically,swaminathan2015counterfactual} are proposed to improve those two methods with different settings. 

As mentioned above, the second challenge is how to deal with confounders. When all the confounders are observed, we can directly optimize the unbiased reward function mentioned in the previous paragraph. However, when there exist unobserved confounders, it can lead to policies that introduce harm rather than benefit, as is generally the case with observational data~\cite{kallus2018confounding}. The confounding-robust policy learning framework is proposed in~\cite{kallus2018confounding}, optimizing the policy over an uncertain set for propensity weights so that the unobserved confounders can be controlled.

\subsection{Other Applications}
The applications of causal inference are not limited to the areas mentioned above, and areas related to effectiveness measurement, decision making, or handling selection bias, are all potential applications.

\textbf{Education}. In the education area, by comparing the outcome of different teaching methods on the student population, a better teaching method can be decided. Moreover, ITE estimation can enhance personalized learning by estimating the outcome of each student on different teaching methods. For example, ITE estimation is developed to answer the questions ``Would this particular student benefit more from the video hint or the text hint when this student cannot solve a problem?'', so that an Intelligent Tutor
System (ITS) can decide which hint is more suitable for a specific student~\cite{zhao2017education}. 

\textbf{Political decision}. In the politics area, causal inference can provide decision support. For example, various methods~\cite{shalit2017estimating, johansson2016learning,yoon2018ganite,yao2018representation,schwab2018perfect} have been developed on the jobs dataset aiming to answer the question``who would benefit most from subsidized job training?''.  Causal inference can also help with political decisions such as whether a policy should be promoted to large population size.

\textbf{Improving Machine learning methods}. In addition to the decision support, various balancing methods that can handle the selection bias ( mentioned in Section~\ref{Section-3}), can also be extended to improve the stability of machine learning methods. In~\cite{kuang2018stable}, the reweighting method is adopted to improve the generalization ability of learned models for unknown environments (i.e., unknown test data). To be specific, the weight of each sample is added to the prediction loss function as a regularization, which is formulated as:
$
\sum_{j = 1}^p||\frac{\phi(X^T_{.,-j}) (R \odot X_{.,j})}{ R^T X_{.,j}} - \frac{\phi(X^T_{.,-j}) (R \odot (1-X_{.,j}))}{ R^T (1-X_{.,j})} ||_2^2
$, where $p$ is the number of total features, $\phi(\cdot)$ is the feature transformation function such as neural network, $X_{.,j}$ is the $j$-th feature in $X$, $X_{.,-j}$ is the features in $X$ except the $j$-th feature, $R\in \mathcal{R}^N$ is the global sample weights with N as the number of total samples. This balancing regularizer extends the CBPS method discussed in~\ref{subsection: sample re-weighting}, by taking the $j$-th feature as the treatment and the remaining features as the background variables, and then combining all the features to obtain the global balancing weight. 

\section{Conclusions}
\label{sec: conclusions}
Causal inference has been an attractive research topic for a long time as it provides an effective way to uncover causal relationships in real-world problems. Nowadays, the flourishing of machine learning brings new vitality into this area, and meanwhile, the incisive ideas in the causal inference area promote the development of machine learning. 
In this survey, we provide a comprehensive review of the methods under the well-known potential outcome framework. As the potential outcome framework relies on the three assumptions, the methods are separated into two categories. One category relies on those assumptions, while the other one relaxes some of the assumptions. For each category, we provide thorough discussions, comparisons, and summarization of the reviewed methods. The available benchmark datasets and open-source codes of those methods are also listed. Finally, some representative real-world applications of causal inference are introduced, such as advertising, recommendation, medicine, and reinforcement learning.

\bibliographystyle{abbrv}
\bibliography{survey}

\end{document}